\documentclass[]{pasj02_orcid}  

\usepackage[dvipsnames]{xcolor}
\usepackage[switch,mathlines]{lineno} 
\usepackage{natbib} 
\definecolor{xlinkcolor}{cmyk}{1,1,0,0}
\usepackage{url}
\usepackage{comment}
\usepackage{lscape}
\usepackage{multirow}
\usepackage{threeparttable}
\usepackage{bm}

\usepackage[colorlinks=true, linkcolor=xlinkcolor, citecolor=xlinkcolor, urlcolor=xlinkcolor]{hyperref}
\makeatletter
\providecommand{\theH@article}{}

\makeatother
\usepackage{CJKutf8}

\jyear{2026}
\Received{}
\Accepted{}

\begin{document} 

\title{
Hidden in Pixels I: Discovery of dual ``little red dots'' indicates excess clustering on kilo-parsec scales
}

\author{
Takumi S. \textsc{Tanaka},\altaffilmark{1,2,3} \email{takumi.tanaka@ipmu.jp} \orcid{0009-0003-4742-7060}
John D. \textsc{Silverman},\altaffilmark{1,2,3,4} \orcid{0000-0002-0000-6977}
Kazuhiro \textsc{Shimasaku},\altaffilmark{2,5} \orcid{0000-0002-2597-2231}
Junya \textsc{Arita},\altaffilmark{2} \orcid{0009-0007-0864-7094}
Hollis B. \textsc{Akins},\altaffilmark{6} \orcid{0000-0003-3596-8794}
Feige \textsc{Wang},\altaffilmark{7} \orcid{0000-0002-7633-431X}
Kohei \textsc{Inayoshi},\altaffilmark{8} \orcid{0000-0001-9840-4959}
Xuheng \textsc{Ding},\altaffilmark{9} \orcid{0000-0001-8917-2148}
Masafusa \textsc{Onoue},\altaffilmark{10,1,8} \orcid{0000-0003-2984-6803}
Zhaoxuan \textsc{Liu},\altaffilmark{1,2,3} \orcid{0000-0002-9252-114X}
Caitlin M. \textsc{Casey},\altaffilmark{11,6,12} \orcid{0000-0002-0930-6466}
Erini \textsc{Lambrides},\altaffilmark{13} \orcid{0000-0003-3216-7190}
Vasily \textsc{Kokorev},\altaffilmark{6,14} \orcid{0000-0002-5588-9156}
Shuowen \textsc{Jin},\altaffilmark{12,15} \orcid{0000-0002-8412-7951}
Andreas L. \textsc{Faisst},\altaffilmark{16} \orcid{0000-0002-9382-9832}
Jianwei \textsc{Lyu},\altaffilmark{17} \orcid{0000-0002-6221-1829}
Jan-Torge \textsc{Schindler},\altaffilmark{18} \orcid{0000-0002-4544-8242}
Yunjing \textsc{Wu},\altaffilmark{1} \orcid{0000-0003-0111-8249}
Nicole \textsc{Drakos},\altaffilmark{19} \orcid{0000-0003-4761-2197}
Yue \textsc{Shen},\altaffilmark{20,21} \orcid{0000-0003-1659-7035}
Junyao \textsc{Li},\altaffilmark{20} \orcid{0000-0002-1605-915X}
Mingyang \textsc{Zhuang},\altaffilmark{20} \orcid{0000-0001-5105-2837}
Qinyue \textsc{Fei},\altaffilmark{22,23} \orcid{0000-0001-7232-5355}
Kei \textsc{Ito},\altaffilmark{12,15} \orcid{0000-0002-9453-0381}
Wei Leong \textsc{Tee},\altaffilmark{17} \orcid{0000-0003-0747-1780}
Weizhe \textsc{Liu} \begin{CJK}{UTF8}{gbsn}(刘伟哲)\end{CJK},\altaffilmark{24} \orcid{0000-0003-3762-7344}
Wenke \textsc{Ren},\altaffilmark{25,26} \orcid{0000-0002-3742-6609}
Tomokazu \textsc{Kiyota},\altaffilmark{27,28} \orcid{0009-0004-4332-9225}
Zi-Jian \textsc{Li},\altaffilmark{29,30} \orcid{0000-0001-7634-1547}
Suin \textsc{Matsui},\altaffilmark{2} \orcid{0009-0009-7403-8603}
Makoto \textsc{Ando},\altaffilmark{28,31} \orcid{0000-0002-4225-4477}
Shun \textsc{Hatano},\altaffilmark{32} \orcid{0000-0002-5816-4660}
Michiko S. \textsc{Fujii},\altaffilmark{2} \orcid{0000-0002-6465-2978}
Jeyhan S. \textsc{Kartaltepe},\altaffilmark{33} \orcid{0000-0001-9187-3605}
Anton M. \textsc{Koekemoer},\altaffilmark{34} \orcid{0000-0002-6610-2048}
Daizhong \textsc{Liu},\altaffilmark{35} \orcid{0000-0001-9773-7479}
Henry Joy \textsc{McCracken},\altaffilmark{36} \orcid{0000-0002-9489-7765}
Jason \textsc{Rhodes},\altaffilmark{37} \orcid{0000-0002-4485-8549}
Brant E. \textsc{Robertson},\altaffilmark{38} \orcid{0000-0002-4271-0364}
Maximilien \textsc{Franco},\altaffilmark{39,6} \orcid{0000-0002-3560-8599}
Koki \textsc{Kakiichi},\altaffilmark{12,40} \orcid{0000-0001-6874-1321}
Jinyi \textsc{Yang},\altaffilmark{41} \orcid{0000-0001-5287-4242}
Romain A. \textsc{Meyer},\altaffilmark{42} \orcid{0000-0001-5492-4522}
Irham T. \textsc{Andika},\altaffilmark{43,44} \orcid{0000-0001-6102-9526}
Aidan P. \textsc{Cloonan},\altaffilmark{45} \orcid{0000-0001-9978-2601}
Xiaohui \textsc{Fan},\altaffilmark{24} \orcid{0000-0003-3310-0131}
Ghassem \textsc{Gozaliasl},\altaffilmark{46,47} \orcid{0000-0002-0236-919X}
Santosh \textsc{Harish},\altaffilmark{48} \orcid{0000-0003-0129-2079}
Christopher C. \textsc{Hayward},\altaffilmark{49} \orcid{0000-0003-4073-3236}
Marc \textsc{Huertas-Company},\altaffilmark{49,50,51,52,53} \orcid{0000-0002-1416-8483}
Darshan \textsc{Kakkad},\altaffilmark{34,54} \orcid{0000-0002-2603-2639}
Tomoya \textsc{Kinugawa},\altaffilmark{55,56,5} \orcid{0000-0002-3033-4576}
Mingyu \textsc{Li},\altaffilmark{57} \orcid{0000-0001-6251-649X}
Namrata \textsc{Roy},\altaffilmark{4} \orcid{0000-0002-4430-8846}
Marko \textsc{Shuntov},\altaffilmark{12,40} \orcid{0000-0002-7087-0701}
Margherita \textsc{Talia},\altaffilmark{58,59} \orcid{0000-0003-4352-2063}
Sune \textsc{Toft},\altaffilmark{12,40} \orcid{0000-0003-3631-7176}
Aswin P. \textsc{Vijayan},\altaffilmark{12,15} \orcid{0000-0002-1905-4194}
Yiyang \textsc{Zhang}\altaffilmark{9} \orcid{0009-0006-8707-023X}
}
\altaffiltext{1}{Kavli Institute for the Physics and Mathematics of the Universe (WPI), The University of Tokyo Institutes for Advanced Study, The University of Tokyo, Kashiwa, Chiba 277-8583, Japan}
\altaffiltext{2}{Department of Astronomy, Graduate School of Science, The University of Tokyo, 7-3-1 Hongo, Bunkyo-ku, Tokyo 113-0033, Japan}
\altaffiltext{3}{Center for Data-Driven Discovery, Kavli IPMU (WPI), UTIAS, The University of Tokyo, Kashiwa, Chiba 277-8583, Japan}
\altaffiltext{4}{Center for Astrophysical Sciences, Department of Physics and Astronomy, Johns Hopkins University, Baltimore, MD 21218, USA}
\altaffiltext{5}{Research Center for the Early Universe, Graduate School of Science, The University of Tokyo, 7-3-1 Hongo, Bunkyo-ku, Tokyo 113-0033, Japan}
\altaffiltext{6}{Department of Astronomy, The University of Texas at Austin, 2515 Speedway Boulevard Stop C1400, Austin, TX 78712, USA}
\altaffiltext{7}{Department of Astronomy, University of Michigan, 1085 S. University Avenue, Ann Arbor, MI 48109, USA}
\altaffiltext{8}{Kavli Institute for Astronomy and Astrophysics, Peking University, Beijing 100871, China}
\altaffiltext{9}{School of Physics and Technology, Wuhan University, Wuhan 430072, China}
\altaffiltext{10}{Waseda Institute for Advanced Study (WIAS), Waseda University, 1-21-1, Nishi-Waseda, Shinjuku, Tokyo 169-0051, Japan; Center for Data Science, Waseda University, 1-6-1, Nishi-Waseda, Shinjuku, Tokyo 169-0051, Japan}
\altaffiltext{11}{Department of Physics, University of California, Santa Barbara, Santa Barbara, CA 93109, USA}
\altaffiltext{12}{Cosmic Dawn Center (DAWN), Denmark}
\altaffiltext{13}{NASA-Goddard Space Flight Center, Code 662, Greenbelt, MD, 20771, USA}
\altaffiltext{14}{Kapteyn Astronomical Institute, University of Groningen, 9700 AV Groningen, The Netherlands}
\altaffiltext{15}{DTU Space, Technical University of Denmark, Elektrovej, Building 328, 2800, Kgs. Lyngby, Denmark}
\altaffiltext{16}{Caltech/IPAC, 1200 E. California Blvd. Pasadena, CA 91125, USA}
\altaffiltext{17}{Steward Observatory, University of Arizona, 933 N Cherry Avenue, Tucson, AZ 85721, USA}
\altaffiltext{18}{Hamburger Sternwarte, University of Hamburg, Gojenbergsweg 112, D-21029 Hamburg, Germany}
\altaffiltext{19}{Department of Physics and Astronomy, University of Hawaii, Hilo, 200 W Kawili St., Hilo, HI 96720, USA}
\altaffiltext{20}{Department of Astronomy, University of Illinois at Urbana-Champaign, Urbana, IL 61801, USA}
\altaffiltext{21}{National Center for Supercomputing Applications, University of Illinois at Urbana-Champaign, Urbana, IL 61801, USA}
\altaffiltext{22}{David A. Dunlap Department of Astronomy and Astrophysics, University of Toronto, 50 St. George Street, Toronto, Ontario, M5S 3H4, Canada}
\altaffiltext{23}{Department of Astronomy, School of Physics, Peking University, Beijing 100871, China}
\altaffiltext{24}{Steward Observatory, University of Arizona, 933 N. Cherry Ave., Tucson, AZ 85721, USA}
\altaffiltext{25}{CAS Key Laboratory for Research in Galaxies and Cosmology, Department of Astronomy, University of Science and Technology of China, Hefei, Anhui 230026, People's Republic of China}
\altaffiltext{26}{School of Astronomy and Space Science, University of Science and Technology of China, Hefei 230026, People's Republic of China}
\altaffiltext{27}{Department of Astronomical Science, The Graduate University for Advanced Studies, SOKENDAI, 2-21-1 Osawa, Mitaka, Tokyo, 181-8588, Japan}
\altaffiltext{28}{National Astronomical Observatory of Japan, 2-21-1 Osawa, Mitaka, Tokyo, 181-8588, Japan}
\altaffiltext{29}{Chinese Academy of Sciences South America Center for Astronomy (CASSACA), National Astronomical Observatories of China (NAOC), CAS, 20A Datun Road, Beijing 100012, China}
\altaffiltext{30}{School of Astronomy and Space Sciences, University of Chinese Academy of Sciences, Beijing 100049, China}
\altaffiltext{31}{Institute for Cosmic Ray Research, The University of Tokyo, 5-1-5 Kashiwanoha, Kashiwa, Chiba 277-8582, Japan}
\altaffiltext{32}{Department of Astronomical Science, SOKENDAI (The Graduate University for Advanced Studies), Osawa 2-21-1, Mitaka, Tokyo, 181-8588, Japan}
\altaffiltext{33}{Laboratory for Multiwavelength Astrophysics, School of Physics and Astronomy, Rochester Institute of Technology, 84 Lomb Memorial Drive, Rochester, NY 14623, USA}
\altaffiltext{34}{Space Telescope Science Institute, 3700 San Martin Drive, Baltimore, MD 21218, USA}
\altaffiltext{35}{Purple Mountain Observatory, Chinese Academy of Sciences, 10 Yuanhua Road, Nanjing 210023, China}
\altaffiltext{36}{Institut d’Astrophysique de Paris, UMR 7095, CNRS, and Sorbonne Université, 98 bis boulevard Arago, 75014 Paris, France}
\altaffiltext{37}{Jet Propulsion Laboratory, California Institute of Technology, 4800 Oak Grove Drive, Pasadena, CA 91001, USA}
\altaffiltext{38}{Department of Astronomy and Astrophysics, University of California, Santa Cruz, 1156 High Street, Santa Cruz, CA 95064, USA}
\altaffiltext{39}{Université Paris-Saclay, Université Paris Cité, CEA, CNRS, AIM, 91191 Gif-sur-Yvette, France}
\altaffiltext{40}{Niels Bohr Institute, University of Copenhagen, Jagtvej 128, DK-2200 Copenhagen N, Denmark}
\altaffiltext{41}{Department of Astronomy, University of Michigan, 1085 S. University Ave., Ann Arbor, MI 48109, USA}
\altaffiltext{42}{Department of Astronomy, University of Geneva, Chemin Pegasi 51, 1290 Versoix, Switzerland}
\altaffiltext{43}{Technical University of Munich, TUM School of Natural Sciences, Department of Physics, James-Franck-Str. 1, 85748 Garching, Germany}
\altaffiltext{44}{Max-Planck-Institut für Astrophysik, Karl-Schwarzschild-Str. 1, 85748 Garching, Germany}
\altaffiltext{45}{Department of Astronomy, University of Massachusetts, 710 North Pleasant Street, Amherst, MA 01003, USA}
\altaffiltext{46}{Department of Computer Science, Aalto University, PO Box 15400, Espoo, FI-00 076, Finland}
\altaffiltext{47}{Department of Physics, Faculty of Science, University of Helsinki, 00014-Helsinki, Finland}
\altaffiltext{48}{Laboratory for Multiwavelength Astrophysics, School of Physics and Astronomy, Rochester Institute of Technology, 84 Lomb Memorial Drive, Rochester, NY14623, USA}
\altaffiltext{49}{Center for Computational Astrophysics, Flatiron Institute, 162 Fifth Avenue, New York, NY 10010, USA}
\altaffiltext{50}{Instituto de Astrofísica de Canarias (IAC), La Laguna 38205, Spain}
\altaffiltext{51}{Observatoire de Paris, LERMA, PSL University, 61 avenue de l’Observatoire, 75014 Paris, France}
\altaffiltext{52}{Université Paris-Cité, 5 rue Thomas Mann, 75014 Paris, France}
\altaffiltext{53}{Universidad de La Laguna, Avda. Astrofísico Fco. Sanchez, La Laguna, Tenerife, Spain}
\altaffiltext{54}{Centre for Astrophysics Research, University of Hertfordshire, Hatfield, AL10 9AB, UK}
\altaffiltext{55}{Faculty of Engineering, Shinshu University, 4-17-1, Wakasato, Nagano-shi, Nagano 380-8553, Japan}
\altaffiltext{56}{Research Center for Aerospace System, Shinshu University, 4-17-1, Wakasato, Nagano-shi, Nagano 380-8553, Japan}
\altaffiltext{57}{Department of Astronomy, Tsinghua University, Beijing 100084, China}
\altaffiltext{58}{University of Bologna, Department of Physics and Astronomy (DIFA), Via Gobetti 93/2, I-40129, Bologna, Italy}
\altaffiltext{59}{INAF – Osservatorio di Astrofisica e Scienza dello Spazio, via Gobetti 93/3 - 40129, Bologna - Italy}


\maketitle
\begin{abstract}
``Little Red Dots'' (LRDs) are an abundant high-redshift population newly discovered by the James Webb Space Telescope (JWST) and considered to be an early growth phase of supermassive black holes (SMBHs). Using a method of pixel-by-pixel color selection and relaxing the compactness criteria, we identify four dual LRD candidates in the COSMOS-Web survey with projected separations of $0.\!\!^{\prime\prime}2\,\mathchar`-\,1.\!\!^{\prime\prime}2$. A comparison between existing LRD samples and mock data reveals that the projected separations of these dual LRD candidates are unlikely to result from chance projections of objects at different redshifts. Furthermore, two of the four systems are covered by COSMOS-3D slitless spectroscopy, and a single-line detection at the same observed wavelength for each LRD in a pair strongly supports that they are at identical redshifts. Assuming that the detected lines are H$\alpha$ based on its high equivalent width and broad profile, the spectroscopic redshifts of $z=5.822$ and $5.464$ for the two pairs are consistent with their photometric redshifts, yielding projected separations of $1.64$ and $7.36\,{\rm kpc}$. These discoveries suggest that the angular auto-correlation function (ACF) of LRDs exhibits an excess ($\sim50$ times) on sub-arcsec (kilo-parsec) separations compared to an extrapolation of a power-law ACF of LRDs measured over $10^{\prime\prime}\,\mathchar`-\,100^{\prime\prime}$. Our sample is likely to represent precursors of mergers between LRDs, and such mergers may be one of the mechanisms that can drive the rapid growth of SMBHs in their early evolutionary stages.
\end{abstract}


\section{Introduction}\label{s:intro}
Supermassive black holes (SMBHs) are a key component in the galaxy evolution scenario.
The tight relation between black hole mass ($M_{\rm BH}$) and galaxy properties, such as stellar velocity dispersion, bulge mass, and stellar mass ($M_*$), observed in the local universe (e.g., \citealt{Magorrian1998, Ferrarese2000, Marconi2003, Haring2004, Gultekin2009, Graham2011, Beifiori2012, Kormendy2013, Reines2015}), implies that SMBHs influence star formation in galaxies through feedback from active galactic nucleus (AGN) activity (AGN feedback, e.g., \citealt{Springel2005, DiMatteo2008, Hopkins2008, Fabian2012, DeGraf2015, Harrison2017}).

The James Webb Space Telescope (JWST), with deep observations at infrared wavelengths, has enabled the discovery of high-$z$ low-luminosity AGNs, thus expanding the parameter space and reducing observational biases in the studies of the evolution of SMBHs and their relationship to their host galaxies.
In addition to searches using photometric selection with NIRCam and MIRI (e.g., \citealt{Onoue2023, Labbe2023a, Furtak2023, Barro2024, Kokorev2024_census, Lyu2024_miri_sel}), deep near-infrared spectroscopic observations have identified numerous type-I (e.g., \citealt{Kocevski2023, Harikane2023_agn, Maiolino2023, Matthee2024}) and type-II (\citealt{Scholtz2023}) AGNs.

One of the surprising discoveries with JWST is a new population, named ``little red dots'' (LRDs, \citealt{Matthee2024}), characterized by compact morphologies and V-shape SEDs: blue excess in the rest-UV wavelengths and red color in the rest-optical wavelengths \citep[e.g.,][]{Labbe2023a, Furtak2024, Matthee2024, Kocevski2023, Akins2023, Barro2024, Akins2024}.
Spectroscopic observations find broad Balmer lines with the full width at half maximum (FWHM) of $\gtrsim1000\,{\rm km\,s^{-1}}$ \citep[e.g.,][]{Labbe2023b, Furtak2024, Matthee2024, Greene2024, Akins2024_spec}, possibly suggesting that LRDs may be type-I AGNs.
The valley of their V-shape SEDs is typically located near the Balmer limit \citep[e.g.,][]{Setton2024}, and LRDs often exhibit a prominent Balmer break \citep[e.g.,][]{deGraaff2025, Kokorev2024_break, Setton2024, Baggen2024} and narrow Balmer absorption features together with the broad Balmer emission lines \citep[e.g.,][]{Matthee2024}.
The non- or tentative detection of X-ray emission \citep{Maiolino2024_chandra, Ananna2024, Yue2024, Akins2024}, infrared emission from a warm torus \citep{Perez2024, Akins2024, Casey2024_lrd}, radio emission \citep{Akins2024, Perger2024, mazzolari2024}, and significant photometric variability \citep{Kokubo2024, Zhang2024_variability, tee2024_variability} suggested that LRDs are not typical AGNs.

Recent studies have instead proposed a BH envelope (BH*) interpretation \citep[e.g.][]{Naidu2025}, where LRDs are rapidly growing BHs surrounded within dense gas envelopes with densities of $n_{\rm H} \gtrsim 10^9\,{\rm cm^{-3}}$ \citep{InayoshiMaiolino2024, Kido2025}, potentially undergoing super-Eddington accretion \citep[e.g.,][]{Greene2024, Lambrides2024, InayoshiMaiolino2024, inayoshi2024, mazzolari2024, Inayoshi2025}.
This model can simultaneously describe the rest-optical red SED steeply rising from around the Balmer limit, the presence of an intense Balmer break, and the lack of detectable variability and of X-ray, hot-torus, and radio emissions.
The rest-optical-to-near-infrared spectra, which can be well described by blackbody emission with effective temperatures ($T_{\rm eff}$) of $\sim 3000\,\mathchar`-\,6000\,{\rm K}$ \citep[e.g.,][]{deGraaff2025_spec_stat, Lin2025_lowz, Umeda2025_BHstar}, together with the presence of metal and ${\rm H_2O}$ absorption features \citep[e.g.,][]{Lin2025_lowz, Wang2026_H2O}, further support this interpretation.

Another important feature of LRDs is their cosmic abundance.
Their number density is 1-2\,dex higher \citep{Kokorev2024_census, Kocevski2024, Akins2024} than an extrapolation from the luminosity functions of high-$z$ luminous quasars \citep{Matsuoka2016, Niida2020}.
The redshift evolution of the number density of LRDs has also been investigated \citep[e.g.,][]{Kocevski2024, Ma2025, Zhuang2025, Tanaka2025_z10LRD}.
Their evolution can be described by a log-normal function of cosmic time that peaks at $z \sim 6$, suggesting a possible connection to early phases of black hole accretion activity \citep{Inayoshi2025}.
In addition, the fraction of LRDs out of the overall galaxy population is reported to increase toward higher redshifts even at $z \gtrsim 6$ \citep{Tanaka2025_z10LRD}.
This trend implies that LRDs may represent a key population for understanding SMBH formation in the early Universe.
To explain the rapid formation of \hbox{high-$z$} SMBHs, several theoretical studies before JWST proposed mechanisms such as super-Eddington accretion and massive seed BHs, including direct collapse BHs \citep[e.g.,][]{Haehnelt1993, Loeb1994, Bromm2003, Begelman2006, Omukai2008, Alexander2014, Bhowmick2022, Zhu2022}.
    
BH mergers represent another possible mechanism for promoting rapid BH growth \citep{Yu2002, Capelo2015, Liu2024_gw}.
An assessment of their impact requires constraining the merger rate through observations and comparing with theoretical studies \citep[e.g.,][]{Liu2024_gw}.
In fact, theoretical efforts \citep[e.g.,][]{Volonteri2022, Chen2023_ASTRID, BM2024} and observational studies \citep[e.g.,][]{Alonso2007, Woods2007, Koss2010, Ellison2011, Silverman2011, Weston2017,Goulding2018, Ellison2019, Perna2023_highnumber, Steffen2023, Bickley2024, Comerford2024, LaMarca2024, ellison2024, Uematsu2024} demonstrate that mergers and interactions trigger AGNs by efficiently supplying fuel to the central regions of galaxies.
Furthermore, accurate measurements of the merger rate at early cosmic epochs will also improve our understanding of the prospects of detecting gravitational waves from distant BH mergers \citep[e.g.,][]{DeRosa2019, Goulding2019, Barausse2020} in future observations, such as the Laser Interferometer Space Antenna (LISA, \citealt{LISA2017}).

Previous studies with JWST have also explored dual AGNs and close pairs of AGNs at the same redshift from 1D spectra \citep{Maiolino2023}, images \citep{Harikane2024, Li2024_dual}, and integral field unit (IFU) spectra \citep{Perna2023_z33, Perna2023_highnumber, Ubler2024, Ishikawa2024, zamora2024}.
While these studies show that there are many merging AGNs in the high-$z$ universe with a fraction of roughly 10-30\% \citep{Li2024_dual, Perna2023_highnumber}, dual LRDs have yet to be discovered\footnote{After the first submission of this manuscript, \citet{Merida2025} and \citet{Yanagisawa2026} reported a photometric candidate of close LRD pairs in gravitationally lensed fields.}.
One possible reason may be that the photometric selection of LRDs in previous studies uses aperture (or total) color and compactness to identify high-$z$ LRD candidates.
The widely used color criteria is $m_{\rm F277W} - m_{\rm F444W} > 1.5$ (e.g., \citealt{Barro2024, Akins2024}) for selecting objects with a red color.
To constrain the compactness, some studies (e.g., \citealt{Greene2024, Akins2024}) use the ratio of aperture fluxes with different diameters. For example, \cite{Akins2024} uses the aperture photometry of F444W with 0\farcs2 and 0\farcs5 diameters and selects compact objects with a flux ratio of $F_{\rm F444W}\left(0\farcs2\right)/F_{\rm F444W}\left(0\farcs5\right) > 0.5$.
These selection methods may overlook LRDs with nearby components since very close companions with projected separation of $\theta\lesssim0\farcs5$ (correspond to $\lesssim2\mathchar`-3\,{\rm kpc}$ at $z\sim5\mathchar`-7$) can lead to an underestimation of the compactness or incorrect assessment of their color (particularly in cases where the companions are bluer than the LRDs; see \citealt{Tanaka2024CB} for an example).
Therefore, this approach would, as intended, favor the selection of isolated ``dot''-like objects and overlook LRDs without nearby components.

We develop a pixel-by-pixel color selection method to address this ``dots'' bias.
As the first paper in a series on this new method, this paper focuses on the first dual LRD candidates discovered from objects selected with the pixel-by-pixel color selection method.
In this paper, we introduce the data and selection methods in section\,\ref{s:data_method}.
In section\,\ref{s:dual}, we introduce the first dual LRD candidates.
Based on these results, sections\,\ref{s:discussion} and \ref{s:discussion2} discuss the implications of this discovery in terms of small-scale clustering and as potential precursors of BH mergers, respectively.
Throughout this paper, the AB magnitude system \citep{Oke1983} was adopted, and we assume a \citet{Kroupa2001} initial mass function and a standard cosmology with $H_0 = 70\,{\rm km\,s^{-1}\,Mpc^{-1}}$, $\Omega_m = 0.30$, and $\Omega_\Lambda=0.70$.
All separations are given in physical (not comoving) units.

\section{Methods}\label{s:data_method}

\subsection{Data}
To conduct a comprehensive search for LRDs, we use the JWST imaging data from COSMOS-Web (\href{https://www.stsci.edu/jwst/science-execution/program-information?id=1727}{GO\,1727}, PI: J. Kartaltepe and C. Casey, \citealt{Casey2023}), which is a JWST Cycle\,1 JWST treasury survey program covering an area of 0.54\,${\rm deg}^2$ with NIRCam (\citealt{Rieke2023}; F115W, F150W, F277W, F444W, see \citealt{Franco2025_CW_NIRCam}) and 0.19 ${\rm deg}^2$ with MIRI (\citealt{Bouchet2015}; F770W, see \citealt{Harish2025_CW_MIRI}).
In addition to COSMOS-Web, to characterize the identified objects, we also use the NIRCam imaging and wide-field slitless spectroscopy (WFSS) data from COSMOS-3D (\href{https://www.stsci.edu/jwst-program-info/program/?program=5893}{GO\,5893}, PI: K. Kakiichi, Kakiichi et al. in preparation), which mainly performed a WFSS survey with the NIRCam/F444W filter over a part ($0.33\,{\rm deg^2}$) of the COSMOS-Web field.

\subsubsection{NIRCam/imaging data}
All available NIRCam imaging data for COSMOS field were reduced using a custom pipeline, following the procedures outlined in \cite{Franco2025_CW_NIRCam} and \cite{Bagley2023}, with additional details provided in Akins et al. (in preparation).
The raw NIRCam images are processed with the JWST Calibration Pipeline version 1.17.1 \citep{jwstpipe_1171}, supplemented by several custom modifications, most notably the subtraction of $1/f$ noise and sky background. 
We use the Calibration Reference Data System (CRDS)\footnote{\url{jwst-crds.stsci.edu}} pmap 1331.

The $1/f$ noise is subtracted via an iterative source-masking and amp-row median subtraction, as described in \cite{Bagley2023}. 
In addition, wisps, scattered-light features affecting several NIRCam short-wavelength (SW) detectors, are removed by rescaling and subtracting the version 3 templates provided by STScI\footnote{\url{https://stsci.app.box.com/s/1bymvf1lkrqbdn9rnkluzqk30e8o2bne}}.
Background subtraction is performed using the same iterative source-masking approach described in \citet{Bagley2023}.
Astrometric calibration is carried out with the JWST/HST alignment tool \citep[JHAT;][]{Rest2023}, using as a reference catalog an HST/F814W mosaic of the field with a pixel scale of $0.\!\!^{\prime\prime}03,{\rm pixel^{-1}}$ \citep{Koekemoer2007}, whose astrometry is tied to Gaia DR3 \citep{GAIA_DR3}.
All NIRCam mosaics are resampled to a pixel scale of $0.\!\!^{\prime\prime}03\,{\rm pixel^{-1}}$.
While the full mosaic included other programs covering the COSMOS field, the imaging data for the four targets we focus in the main text are purely based on COSMOS-Web and COSMOS-3D. 

\subsubsection{MIRI/imaging data}
The MIRI/F770W imaging data was reduced by the COSMOS-Web team following the procedure presented in \cite{Harish2025_CW_MIRI}.
Some key steps of the processing are briefly summarized below.
The F770W data were processed with the JWST Calibration Pipeline version 1.12.5 \citep{jwstpipe_1125} with JWST Calibration Reference Data System Pipeline Context version 1130.
The reduction includes the standard rate calculation, jump detection, slope fitting, non-linearity correction, flat fielding, flux scale conversion, etc., as well as a highly customized master background subtraction (see \citealt{Harish2025_CW_MIRI}).
Astrometric refinement is likewise performed using a customized procedure, involving source extraction with \texttt{SExtractor} \citep{Bertin1996} and image alignment via the JWST pipeline \texttt{tweakreg} module, anchored to the same absolute reference catalog used for the NIRCam reductions.
To maximize the effective field of view, the Lyot detector area was also processed together with the main imager area, which provides good quality of data after applying our customized master background subtraction.
The F770W mosaic is also resampled to a pixel scale of $0.\!\!^{\prime\prime}03\,{\rm pixel^{-1}}$.

\subsubsection{HST/imaging data}
For the HST data, we use the ACS/F814W mosaics \citep{Koekemoer2007}, which have been reprocessed with up-to-date calibration reference files and aligned to Gaia-DR3 (\!\!\citealt{GAIA_DR3}), providing the astrometric reference frame for the JWST imaging data.
The F814W mosaics are also resampled to a pixel scale of $0.\!\!^{\prime\prime}03\,{\rm pixel^{-1}}$.

\subsubsection{NIRCam/grism data}
The COSMOS-3D NIRCam/WFSS (F444W) observations covered two of the four selected dual LRD candidates described later, and we use the grism spectra to spectroscopically confirm their redshifts and nature.
We use the COSMOS-3D grism data processed with \texttt{unfold\_jwst} version 0.9.0 (Wang et al. in prep.).
The data reduction follows the procedures described in \citep{Wang2023}, using the JWST Calibration Pipeline version 1.17.1 \citep{jwstpipe_1171} together with custom scripts implemented in \texttt{unfold\_jwst}.
We use the CRDS\footnote{\url{jwst-crds.stsci.edu}} pmap 1321.
Note that the COSMOS-3D grism data is also processed with \texttt{grizli}\footnote{\url{https://github.com/gbrammer/grizli}} \citep{grizli_2023}, to reduce the software dependencies and obtain robust results (see \citealt{Meyer2025_OIII} for the details of the processing).
While we use the data processed with \texttt{unfold\_jwst} in this paper, we also confirmed that the results do not change significantly when using the data processed with \texttt{grizli}.

\subsection{Identification of dual LRD candidates}\label{ss:sample_Selection}
\begin{figure*}
 \begin{center}
  \includegraphics[width=14cm]{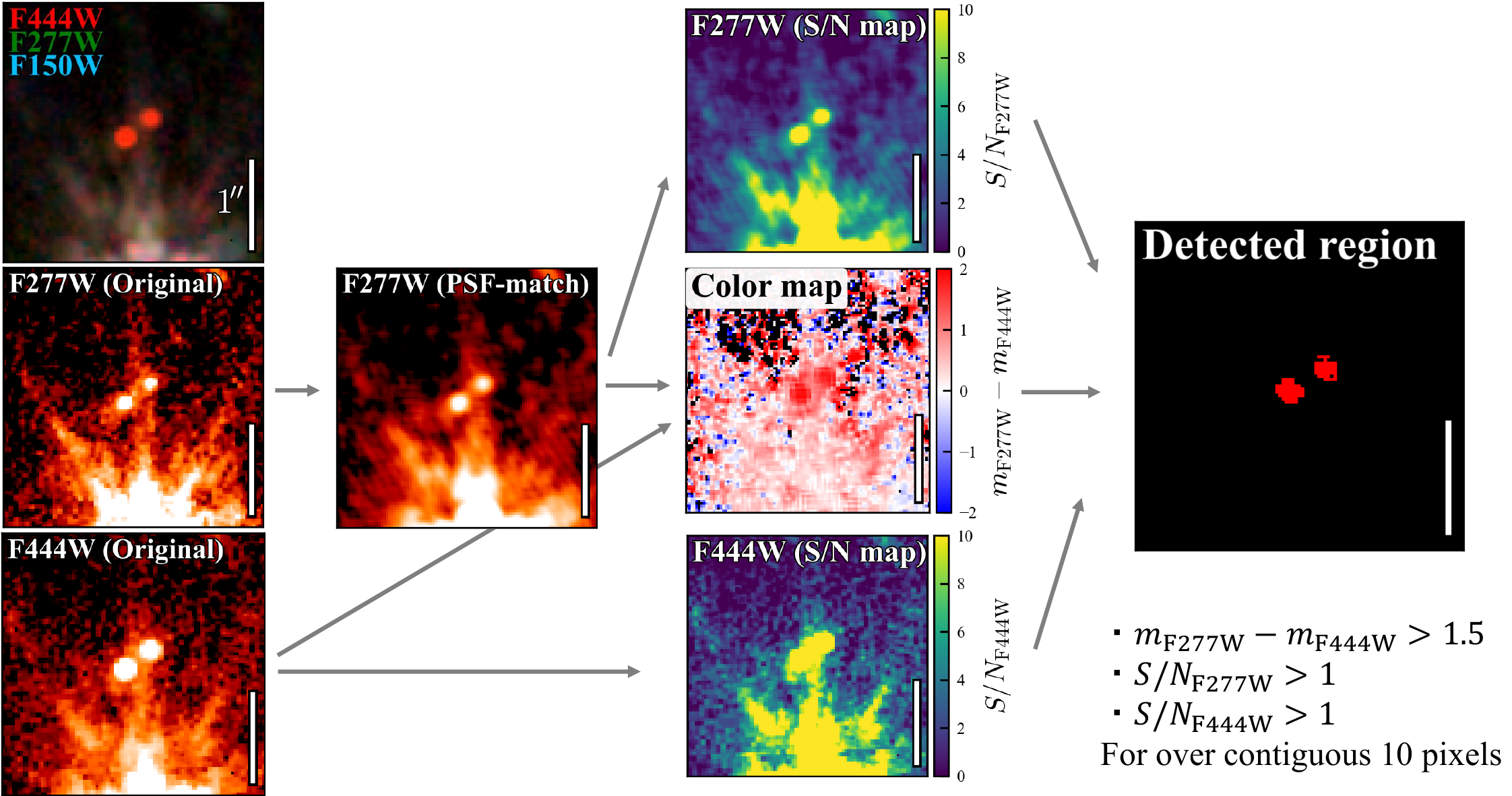} \end{center}
\caption{
Schematic figure of the pixel-by-pixel color selection method. 
We construct $S/N$ and color maps using PSF-matched F277W and F444W images. 
By combining these maps, we identify regions consisting of more than 10 contiguous pixels that satisfy $(S/N)_{\rm pixel}>1.0$ and $(m_{\rm F277W}-m_{\rm F444W})_{\rm pixel}>1.5$. 
This procedure enables us to select regions with LRD-like colors relatively free from contamination.
}
\label{fig:method_schematic}
\end{figure*}

We apply a color selection to each pixel (i.e., a pixel-by-pixel color selection method) instead of using aperture or total photometry.
The full details of the selection method and subsequent catalog will be presented in Paper II (Tanaka et al. in preparation).
Here, we briefly describe the selection method.

First, we perform a pre-selection to limit the parent sample to objects with a high signal-to-noise ratio (S/N) for a robust color estimate in each pixel.
We run {\tt SExtractor} \citep{Bertin1996} on the NIRCam F444W images to perform source detection and simple aperture photometry with a diameter of $0.\!\!^{\prime\prime}5$.
Based on the photometric catalog, we limit our sample to objects detected in F444W with $S/N>10$, following \cite{Greene2024}.

Then, to assess each pixel color, we generate a point spread function (PSF) modeled with {\tt PSFEx} \citep{Bertin2011} for each filter and convolute the images to match the PSF of F444W, which has the largest PSF of the used NIRCam imaging data.
The convolution kernel is generated with {\tt create\_matching\_kernel} function of {\tt photutils.psf.matching} using {\tt HanningWindow} for a window function.
We then create cutout images of $101\,{\rm pixel} \times 101\,{\rm pixel}$ square ($3.\!\!^{\prime\prime}0 \times 3.\!\!^{\prime\prime}0$) for the pre-selected objects.

Next, we generate a $S/N$ map for each filter and a $m_{\rm F277W}-m_{\rm F444W}$ color map for each cutout image (see figure\,\ref{fig:method_schematic}).
From the $S/N$ and color maps, we generate a mask image to extract the region having a color of $\left(m_{\rm F277W}-m_{\rm F444W}\right)_{\rm pixel}>1.5$ (e.g., \citealt{Barro2024, Akins2024}) and a $S/N$ in each pixel is $\left(S/N\right)_{\rm pixel}>1.0$ for both the F277W and F444W images.
Then, if the masked pixels form at least ten contiguous pixels, we select the object as having an LRD-like-colored region.
Utilizing machine-learning-based dimensionality reduction with DINOv2 \citep{DINOv2} and Uniform Manifold Approximation and Projection for Dimension Reduction \citep[UMAP,][]{UMAP} and clustering analysis with Density-Based Spatial Clustering of Applications with Noise (DBSCAN) algorithm, we efficiently removed artifact objects, such as PSF spikes and saturated point sources.

\begin{figure*}
 \begin{center}
  \includegraphics[width=17.8cm]{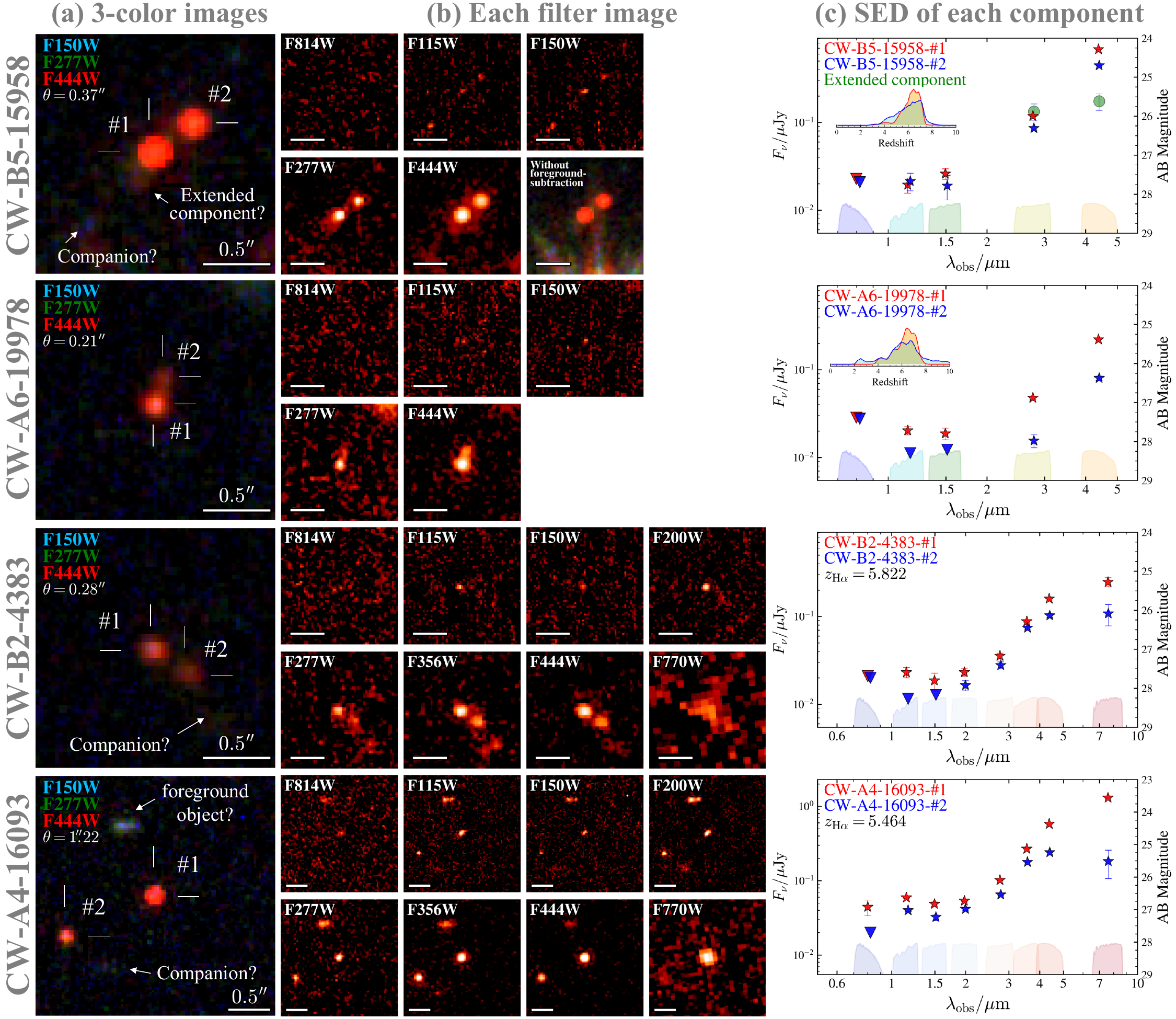} 
 \end{center}
\caption{
(a) Three-color (F444W, F277W, and F150W for RGB) image of each dual LRD candidate. 
We have not matched the PSF between each filter.
For CW-B5-15958, we plot the images after subtracting the foreground type-1 AGN (CID-643; see section\,\ref{ss:cw-b5-15958}).
(b) Individual HST (ACS: F814W) and JWST images (NIRCam: F115W, F150W, F200W, F277W, F356W, and F444W, MIRI: F770W).
For CW-B5-15958, we also display the original three-color image before subtracting the foreground type-1 AGN.
(c) SED of each component estimated from an image-based modeling analysis.
The inverted triangles indicate the $3\sigma$ upper limits due to non-detections in specific filters.
The best-fit model SEDs are shown in figure\,\ref{fig:sed_fitting}.
We also plot probability distributions of $z_{\rm photo}$ for systems without COSMOS-3D spectroscopic confirmation.
}
\label{fig:3col_sed}
\end{figure*}

Through visual inspection of the artifact-cleaned sample, we identified four dual LRD candidates.
Each candidate consists of two compact sources that are suggested to be PSF-like based on morphological fitting (see section\,\ref{sss:morphology}), with colors consistent with those of LRDs within a $3^{\prime\prime}\times3^{\prime\prime}$ cutout, separated by $\sim 0.\!\!^{\prime\prime}2 \,\mathchar`-\, 1.\!\!^{\prime\prime}2$ as shown in figure\,\ref{fig:3col_sed}\,(a).
Throughout the paper, the brighter LRD in each pair is referred to as component \#1 while the fainter LRD is referred to as component \#2.
We also identified many systems in which an LRD is located in close proximity to other types of objects.
Unlike those LRD-galaxy systems, the four dual LRD candidates selected above are distinguished by the fact that not only the primary LRD but also the accompanying companion independently satisfies the same selection criteria (also see figure\,\ref{fig:method_schematic}).
Note that we do not find any systems containing more than two LRDs.

\subsection{Analysis}
\subsubsection{Morphological modeling}\label{sss:morphology}

To confirm their compact morphology and measure the photometry of each LRD within these systems, we perform image-based morphological modeling using {\tt galight} \citep{Ding2020_HST}.
We simultaneously fit each component of the pairs with either a Sérsic profile \citep{Sersic1963, sersic1968} or a PSF to check whether spatially unresolved components describe these candidates or not.
Using the fitting results with four different models, Sérsic-Sérsic, Sérsic-PSF, PSF-Sérsic, and PSF-PSF for the pair of LRD candidates, we determine whether each component is best represented by a PSF or an extended Sérsic profile using the Bayesian Information Criterion (BIC). 
In this study, we use ``significantly lower'' BIC for cases where the BIC is lower by more than 10 compared to the other models.
The center position of each component is fixed based on the PSF-PSF fitting results for F444W.
For F115W and F150W of CW-A6-19978 and CW-B2-4383 and F814W of CW-A4-16093, we also test a case where we do not apply a model to component \#2, i.e., undetected in these filters.
When fitting with Sérsic components, we set the $r_e$ and $n$ ranges of $r_e\in[0.\!\!^{\prime\prime}03, 1.\!\!^{\prime\prime}0]$ and $n\in[0.5, 5.0]$, respectively, to separate the extended components from PSF components effectively.
Using the best model chosen based on the BIC, we measure the photometry of each component as summarized in tables\,\ref{tab:dual_LRD} and \ref{tab:dual_LRD_spec}.

\subsubsection{SED fitting with a LRD model}
To estimate photometric redshifts ($z_{\rm photo}$), we perform SED fitting analysis using a flexible LRD SED model.
Following recent empirical and theoretical understanding of LRDs \citep[e.g.,][]{Naidu2025, Inayoshi2025_seduni}, our model is mainly a superposition of two components: a galaxy component and a BH envelope component dominating the rest-UV and optical part, respectively.

The galaxy component $L_{\rm \lambda, gal}\left(\lambda\right)$ is modeled with Flexible Stellar Population Synthesis \citep[FSPS,][]{Conroy2009, Conroy2010}.
We adopt an exponentially declining delayed-$\tau$ star-formation history (SFH),
\begin{equation}
    {\rm SFR}\left(t\right) \propto t\exp\left(-\frac{t}{\tau}\right),
\end{equation}
where $\tau$ is the e-folding timescale.
For simplicity, we fix a metallicity to solar value, apply \cite{Calzetti2000} simple dust attenuation law, and assume a \cite{Kroupa2001} initial mass function.
The overall galaxy component is scaled by $\log M_*$.

For the BH envelope component, we first model a pure blackbody component $L_{\rm \lambda, BB}\left(\lambda\right)$ following the Planck function normalized by the bolometric luminosity $L_{\rm BB}$ as,
\begin{equation}
    L_{\rm \lambda, BB}\left(\lambda, T_{\rm eff}, L_{\rm BB}\right) = L_{\rm BB}\frac{\pi B_\lambda\left(\lambda, T_{\rm eff}\right)}{\sigma_{\rm SB} T_{\rm eff}^4},
\end{equation}
where $B_\lambda$ and $\sigma_{\rm SB}$ are the Planck specific intensity and the Stefan-Boltzmann's constant.
Then, to model the BH envelope component, we apply the Balmer absorption $A_{\rm Balmer}\left(\lambda\right)$ modeled following the way in \cite{Inayoshi2025_binary} and \cite{Inayoshi2025_seduni} to the blackbody component as $L_{\rm \lambda, BHE}\left(\lambda\right) = A_{\rm Balmer}\left(\lambda\right) \times L_{\rm \lambda, BB}\left(\lambda\right)$.

${\rm H\alpha}$ rest-frame equivalent width (EW) for LRDs can reach ${\rm EW}\sim 2\times10^3\,\text{\AA}$ \citep[e.g.,][]{Maiolino2024_chandra, deGraaff2025_spec_stat}, which can significantly affect their photometry \citep[e.g.,][]{Hviding2025}.
Thus, we also include an ${\rm H\alpha}$ emission line using a rest-frame EW as a free parameter on top of the composite continuum of $L_{\rm \lambda, gal}\left(\lambda\right) + L_{\rm \lambda, BHE}\left(\lambda\right)$
The integrated line luminosity $L_{\rm H\alpha}$ is defined as:
\begin{equation}
    L_{\rm H\alpha} = {\rm EW_{H\alpha}} \times \left\{L_{\rm \lambda, gal}\left(\lambda_{\rm H\alpha}\right) + L_{\rm \lambda, BHE}\left(\lambda_{\rm H\alpha}\right)\right\}.
\end{equation}
Then, the line is added as a Gaussian centered at $\lambda_{\rm H\alpha} = 6563\,\text{\AA}$ with a fixed width of ${\rm FWHM}=1000\,{\rm km\,s^{-1}}$.
Note that the assumption of the fixed FWHM does not affect the results since we only fit the broadband photometric data.

We fit the input photometry using the galaxy + BH envelope + ${\rm H\alpha}$ composite SED model with a Markov Chain Monte Carlo (MCMC) technique.
We assume the uniform priors for the free parameters as follows:
\begin{align*}
    \tau &\in [0.01,30]\,{\rm Gyr},\\
    A_V &\in [0,2]\,{\rm mag},\\
    \log \left(M_*/M_\odot\right) &\in [5,11],\\
    t_{\rm age} &\in [10^{-3}, 0.97\times t_{\rm univ}(z)]\,{\rm Gyr},\\
    T_{\rm eff} &\in [2000,10000]\,{\rm K},\\
    \log\left(L_{\rm BB}/{\rm erg\,s^{-1}}\right) & \in [38,48],\\
    {\rm EW}_{\rm H\alpha,rest} &\in [0,3000]\,\text{\AA},\\
    z_{\rm photo} &\in [0,15],
\end{align*}
where $t_{\rm univ}(z)$ is the age of the Universe at each redshift.
In the fitting, the likelihood is evaluated in flux-density space assuming independent Gaussian uncertainties for each photometric data point,
\begin{equation}
\ln \mathcal{L} = -\frac{1}{2} \sum_i
\left[
\frac{\left(f_{\rm \nu,i}^{\rm obs} - f_{\rm \nu,i}^{\rm model}\right)^2}{\sigma_{\rm \nu,i}^2}
\right],
\end{equation}
where $f_{\rm \nu,i}^{\rm obs}$ and $\sigma_{\rm \nu,i}$ are the observed flux density and its uncertainty in the i-th band, and $f_{\rm \nu,i}^{\rm model}$ is the model-predicted flux density for the i-th band.

In this model, the UV component is reproduced by the galaxy component.
The physical origin of the UV component in LRDs is still under debate, and it remains unclear whether the host galaxy is responsible for the entire UV emission \citep[e.g.,][]{Zhang2025_host}.
Nevertheless, because our SED fitting uses only photometric data as input, we do not discuss the galaxy-related parameters inferred from this fit, such as $M_*$, $A_V$, and SFH.

As described later, an emission line is detected in the grism spectra for CW-B2-4383 and CW-A4-16093.
Assuming that the detected line is ${\rm H\alpha}$, we fix the redshift to the corresponding spectroscopic redshift $z_{\rm H\alpha}$.
However, we do not fix the H$\alpha$ EW, because the limited depth of the data may prevent the detection of a broad component, which could lead to an underestimation of the EW (see section\,\ref{s:dual}).

We also perform an SED-fitting analysis using a flexible type-I AGN template capable of reproducing the SEDs of LRDs, following the methodology of \cite{Akins2024}.
As a result, we obtain broadly consistent photometric redshifts with the BH envelope model described above.
On the other hand, in the type-I AGN model, the steep red spectral slope is reproduced by applying strong dust attenuation to the type-I AGN template.
This requires an attenuation of $A_V\sim3$, and consequently the inferred bolometric luminosity becomes more than two orders of magnitude larger than that obtained with the BH-envelope model described above.
Such strong dust attenuation has been ruled out based on the non-detection of dust re-emission with facilities such as MIRI and ALMA \citep[e.g.,][]{Setton2025, Casey2025_dust}.
Therefore, in this study, we use the results obtained with the BH-envelope model.

\subsubsection{SED fitting with BD templates}
In addition to SED fitting with the LRD model, we also use brown dwarfs (BDs) templates to remove their potential contamination following the method in \cite{Akins2024}.
We construct a comprehensive grid of models spanning a broad parameter space.
We include three models: cloudy-atmosphere models in chemical equilibrium (Sonora-Diamondback; $T\sim900\,\mathchar`-\,2400\,{\rm K}$; \citealt{Morely2024}), cloud-free models in chemical equilibrium (Sonora-Bobcat; $T\sim200\,\mathchar`-\,1300\,{\rm K}$; \citealt{Marley2021}) and disequilibrium (ElfOwl; $T\sim300\,\mathchar`-\,1000\,{\rm K}$; \citealt{Mukherjee2024}), and low-metallicity models (LOWZ; $\left[{\rm M/H}\right] = -1$; \citealt{Meisner2021}). 
The grid explores a range of temperatures ($T=200\,\mathchar`-\,2400\,{\rm K}$), surface gravities ($g\sim\,100\mathchar`-\,3160\,{\rm cm\,s^{-2}}$), and metallicities ($\left[{\rm M/H}\right] =$ $-1$, $-0.5$, and $0$).
We compare the $\chi^2$ of the best-fit BD templates with $\chi^2$ of the best-fit LRD model.

\subsubsection{Spectral line fitting}\label{sss:line_fit}
The COSMOS-3D grism spectra are available for CW-B2-4383 and CW-A4-16093.
For each component in each dual LRD candidate, we extract one-dimensional spectra using an optimal extraction method.
We detect a line-like feature at the same wavelength in both components and perform line fitting using both single- and double-Gaussian models.
In the double-Gaussian fitting, the line width of the narrow component is constrained to $\mathrm{FWHM} < 500\ \mathrm{km\ s^{-1}}$, while the line width of the broad component is constrained to $\mathrm{FWHM} > 500\ \mathrm{km\ s^{-1}}$.
By comparing BIC values of the two models, we assess the presence of a broad component in the emission line.
Note that we do not consider a line spread function (LSF) in this study. 
However, according to the JWST User Document (JDox\footnote{\url{https://jwst-docs.stsci.edu/jwst-near-infrared-camera/nircam-instrumentation/nircam-grisms}}) the spectral resolution of the NIRCam/F444W slitless spectroscopy is $R\sim1600\,\mathchar`-\,1700$, corresponding to a velocity resolution of $\sim 180\,{\rm km\,s^{-1}}$. 
In addition, since the targets discussed in this study are compact objects, the effect of overlap between the spatial and dispersion directions is negligible.
Therefore, LSF does not significantly affect our discussion of broad lines with typical velocities exceeding $1000\,{\rm km\,s^{-1}}$.

\section{Results: dual LRD candidates}\label{s:dual}

\begin{figure*}
 \begin{center}
  \includegraphics[width=15cm]{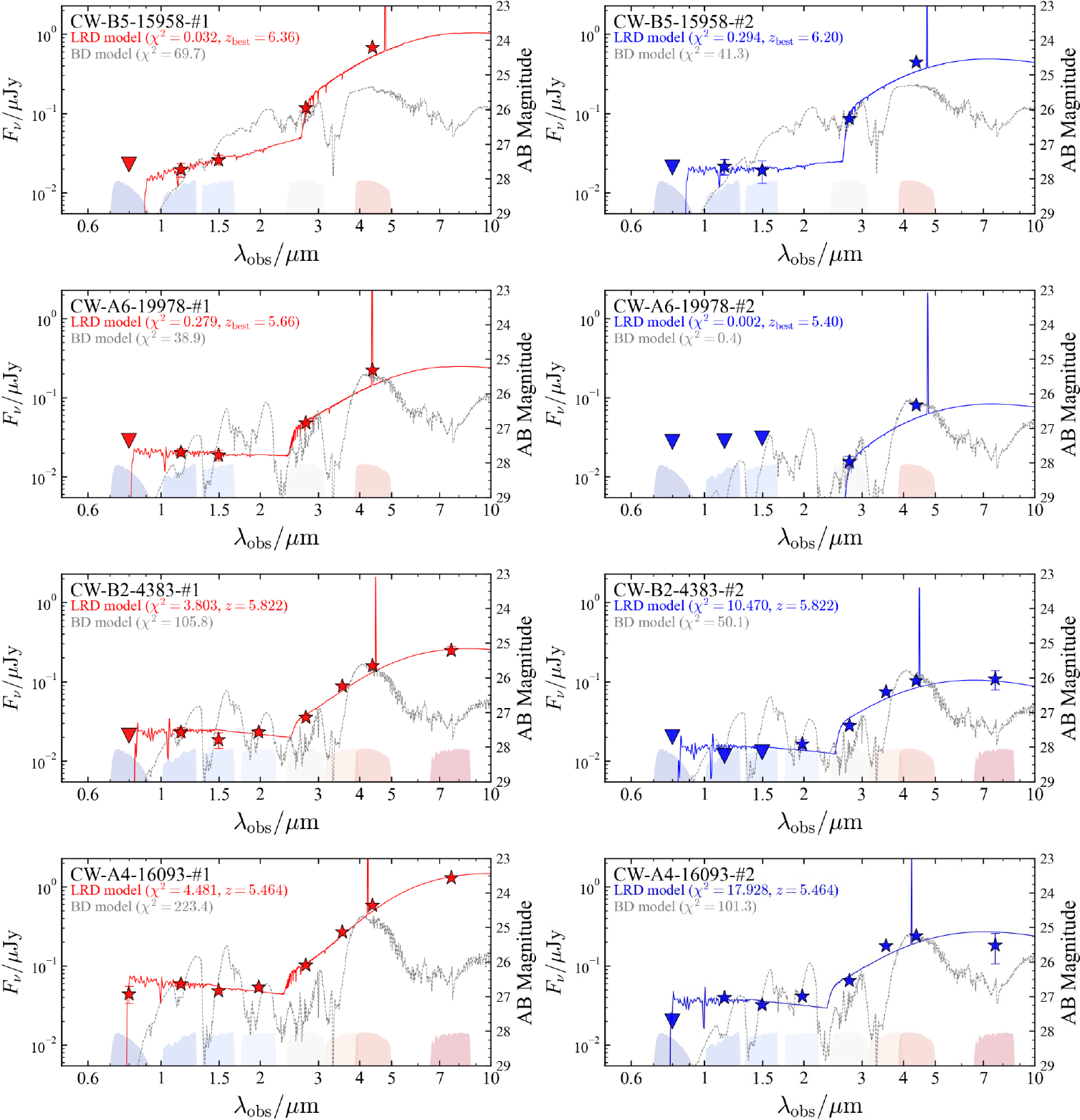} 
 \end{center}
\caption{
SED fitting analysis of the LRD candidates.
The inverted triangle symbols indicate the $3\sigma$ upper limit due to the non-detection in each filter.
Solid colored and dashed gray lines indicate the best-fit model SEDs with the LRD models and the BD templates. 
The
$\chi^2$ and the best-fit $z_{\rm photo}$ (or $z_{\rm H\alpha}$ for the spectroscopically confirmed object) are shown in the upper left corner.
}
\label{fig:sed_fitting}
\end{figure*}

The final inferred parameters from our analysis are summarized in table\,\ref{tab:dual_LRD} and \ref{tab:dual_LRD_spec}.
The best-fit model SEDs from LRD model and BD templates are shown in solid and dashed lines in figure\,\ref{fig:sed_fitting}.
The $\chi^2$ with the LRD model are significantly lower than those for the BD fits, and the BD templates are inconsistent with the observed photometry.
Therefore, we conclude that these objects are not BDs.
Detailed results on each candidate are described in the following subsections.

\begin{table*}[]
\centering
\begin{threeparttable}
\caption{Characteristics of CW-B5-15958 and CW-A6-19978, those without COSMOS-3D spectrum.
}\label{tab:dual_LRD}
\begin{tabular}{c | lll | ll}
& \multicolumn{3}{c|}{CW-B5-15958} & \multicolumn{2}{c}{CW-A6-19978} \\
& \#1 & \#2 & extended component & \#1 & \#2\\
\hline\hline
R.A. & \timeform{10h01m59s.842} & \timeform{10h01m59s.829} & \timeform{10h01m59s.849} & \timeform{09h58m59s.552} & \timeform{09h58m59s.552} \\
Decl. & \timeform{2D26'43''.35} & \timeform{2D26'43''.66} & \timeform{2D26'43''.20} & \timeform{2D07'29''.32} & \timeform{2D07'29''.53} \\
$F_{\rm F814W}/{\rm nJy}$\tnote{a} & $<23$ & $<21$ & $<21$ & $<28$ & $<27$ \\
$F_{\rm F115W}/{\rm nJy}$\tnote{a} & $19.5\pm3.9$ & $21.5\pm4.8$ & $<13$ & $20.3\pm2.2$ & $<11.2$ \\
$F_{\rm F150W}/{\rm nJy}$\tnote{a} & $25.9\pm3.6$ & $19.0\pm6.0$ & $<14$ & $18.8\pm2.9$ & $<12.3$ \\
$F_{\rm F277W}/{\rm nJy}$ & $117.9\pm1.9$ & $85.9\pm3.6$ & $133\pm29$ & $47.7\pm1.4$ & $15.5\pm2.6$ \\
$F_{\rm F444W}/{\rm nJy}$ & $680.1\pm3.1$ & $445.7\pm6.6$ & $174\pm37$ & $221.8\pm2.8$ & $80.2\pm4.3$ \\
$\theta_{\rm F277W}/{\rm arcsec}$ & \multicolumn{2}{c}{0.37} & \multicolumn{1}{c|}{0.20} & \multicolumn{2}{c}{0.21} \\
\hline
$z_{\rm photo}$ & $6.30_{-0.84}^{+0.59}$ & $6.15_{-1.27}^{+0.84}$ & - & $6.36_{-1.08}^{+0.68}$ & $6.23_{-1.43}^{+1.43}$ \\
$z_{\rm system}$\tnote{b} & \multicolumn{2}{c}{$6.41_{-0.70}^{+0.51}$} & - & \multicolumn{2}{c}{$6.42_{-0.63}^{+0.52}$} \\
$d_{\rm sep}\left(z_{\rm system}\right)/{\rm kpc}$ & \multicolumn{2}{c}{$2.03^{+0.14}_{-0.08}$} & \multicolumn{1}{c|}{$1.03_{-0.04}^{+0.10}$} & \multicolumn{2}{c}{$1.15_{-0.05}^{+0.08}$} \\
$T_{\rm eff}/{\rm K}$ & $\left(4.4^{+2.4}_{-1.8}\right)\times10^3$ & $\left(4.2^{+2.9}_{-1.6}\right)\times10^3$ & - & $\left(4.4^{+2.7}_{-1.7}\right)\times10^3$ & $\left(5.1^{+2.9}_{-2.2}\right)\times10^3$ \\
$\log\left(L_{\rm BB}/{\rm erg\,s^{-1}}\right)$ & $44.38_{-0.41}^{+0.27}$ & $44.13_{-0.35}^{+0.27}$ & - & $43.81_{-0.39}^{+0.26}$ & $43.42_{-0.37}^{+0.39}$ \\
$\log\left(M_{\rm BH, BB}/M_\odot\right)$\tnote{c} & $6.28_{-0.41}^{+0.27}$ & $6.03_{-0.35}^{+0.27}$ & - & $5.71_{-0.39}^{+0.26}$ & $5.31_{-0.37}^{+0.39}$ \\
\hline
\end{tabular}
\begin{tablenotes}
\footnotesize
\item[a] Upper limits are based on the $3\sigma$ noise level.
\item[b] The system photometric redshift estimated by simply multiplying $z_{\rm photo}$ probability distribution function (PDF) of each component.
\item[c] Black hole masses estimated from the fitted blackbody luminosity assuming Eddington-limited accretion.
\end{tablenotes}
\end{threeparttable}
\end{table*}

\begin{table*}[]
\centering
\begin{threeparttable}
\caption{Characteristics of CW-B2-4383 and CW-A4-16093, those with COSMOS-3D spectrum.
}\label{tab:dual_LRD_spec}
\begin{tabular}{c | ll | ll}
& \multicolumn{2}{c|}{CW-B2-4383} & \multicolumn{2}{c}{CW-A4-16093}\\
& \#1 & \#2 & \#1 & \#2\\
\hline\hline
R.A. & \timeform{10h00m13s.530} & \timeform{10h00m13s.510} & \timeform{10h01m04s.172} & \timeform{10h01m04s.230} \\
Decl. & \timeform{2D28'18''.76} & \timeform{2D28'18''.68} & \timeform{2D05'29''.85} & \timeform{2D05'28''.98} \\
$F_{\rm F814W}/{\rm nJy}$\tnote{a} & $<21$ & $<20$ & $44\pm10$ & $<20$ \\
$F_{\rm F115W}/{\rm nJy}$\tnote{a} & $23.1\pm3.1$ & $<12$ & $59.3\pm1.6$ & $39.8\pm1.6$ \\
$F_{\rm F150W}/{\rm nJy}$\tnote{a} & $18.6\pm4.1$ & $<12.8$ & $48.6\pm3.0$ & $32.4\pm3.0$ \\
$F_{\rm F200W}/{\rm nJy}$ & $23.2\pm2.2$ & $16.4\pm2.1$ & $53.7\pm1.6$ & $41.6\pm1.6$\\
$F_{\rm F277W}/{\rm nJy}$ & $35.5\pm2.7$ & $27.8\pm2.6$ & $101.8\pm2.1$ & $64.8\pm2.1$ \\
$F_{\rm F356W}/{\rm nJy}$ & $87.6\pm2.8$ & $74.3\pm2.8$ & $267.7\pm1.8$ & $178.4\pm1.8$ \\
$F_{\rm F444W}/{\rm nJy}$ & $159.5\pm4.8$ & $112.2\pm4.9$ & $579.4\pm3.5$ & $240.7\pm3.5$\\
$F_{\rm F770W}/{\rm nJy}$ & $247\pm29$ & $108\pm29$ & $1301\pm15$ & $181\pm75$ \\
$\theta_{\rm F277W}/{\rm arcsec}$ & \multicolumn{2}{c|}{0.31} & \multicolumn{2}{c}{1.23}\\
\hline
$z_{\rm photo}$ & $5.66_{-0.35}^{+0.42}$ & $6.77_{-0.26}^{+0.28}$ & $5.09_{-0.37}^{+0.45}$ & $5.83 _{-0.27}^{+0.28}$ \\
$z_{\rm system}$\tnote{b} & \multicolumn{2}{c|}{-} & \multicolumn{2}{c}{$5.61_{-0.31}^{+0.30}$} \\
$d_{\rm sep}\left(z_{\rm H\alpha}\right)/{\rm kpc}$ & \multicolumn{2}{c|}{$1.64$} & \multicolumn{2}{c}{$7.36$} \\
${\rm EW}_{\rm H\alpha}/\text{\AA}$\tnote{c} & $\left(6.1_{-3.0}^{+4.0}\right)\times10^2$ & $\left(7.5_{-4.9}^{+7.5}\right)\times10^2$ & $\left(4.8_{-2.7}^{+3.4}\right)\times10^2$ & $\left(9.3_{-4.7}^{+6.3}\right)\times10^2$ \\
$T_{\rm eff}/{\rm K}$\tnote{c} & $\left(3.78_{-0.33}^{+0.28}\right)\times10^3$ & $\left(5.4_{-1.9}^{+2.9}\right)\times10^3$ & $\left(3.22_{-0.18}^{+0.16}\right)\times10^3$ & $\left(4.29_{-0.46}^{+0.62}\right)\times10^3$ \\
$\log\left(L_{\rm BB}/{\rm erg\,s^{-1}}\right)$\tnote{c} & $43.69_{-0.08}^{+0.06}$ & $43.47_{-0.67}^{+0.21}$ & $44.40_{-0.05}^{+0.04}$ & $43.60_{-0.23}^{+0.14}$ \\
$\log\left(M_{\rm BH, BB}/M_\odot\right)$\tnote{d} & $5.59_{-0.08}^{+0.06}$ & $5.37_{-0.67}^{+0.21}$ & $6.30_{-0.05}^{+0.04}$ & $5.50_{-0.23}^{+0.14}$ \\
\hline
$\mu/\text{\AA}$ & $44772.8^{+4.7}_{-4.0}$ & $44771.1^{+10.9}_{-9.7}$ & $42421.0^{+2.2}_{-2.2}$ & $42427.9^{+4.4}_{-5.3}$ \\
$z_{\rm H\alpha}$ & $5.8222^{+0.0007}_{-0.0006}$ & $5.8219^{+0.0017}_{-0.0015}$ & $5.4639^{+0.0003}_{-0.0003}$ & $5.4649^{+0.0007}_{-0.0008}$ \\
$\Delta v_{\rm LOS}/{\rm km\,s^{-1}}$ & \multicolumn{2}{c|}{$13^{+76}_{-79}$} & \multicolumn{2}{c}{$47^{+34}_{-40}$} \\
$L_{\rm single}/{\rm 10^{42}\,erg\,s^{-1}}$ & $3.26^{+0.61}_{-0.57}$ & $1.56^{+0.43}_{-0.36}$ & $3.80^{+0.89}_{-0.83}$ & $1.54^{+0.33}_{-0.29}$ \\
${\rm FWHM}_{\rm single}/{\rm km\,s^{-1}}$ & $1118^{+265}_{-224}$ & $625^{+292}_{-172}$ & $1123^{+344}_{-306}$ & $348^{+96}_{-77}$ \\
$L_{\rm narrow}/10^{42}\,{\rm erg\,s^{-1}}$ & $0.59^{+0.31}_{-0.27}$ & - & $0.89^{+0.26}_{-0.23}$ & - \\
$L_{\rm broad}/10^{42}\,{\rm erg\,s^{-1}}$ & $3.66^{+0.83}_{-0.71}$ & - & $4.25^{+1.5}_{-1.1}$ & - \\
${\rm FWHM}_{\rm broad}/{\rm km\,s^{-1}}$ & $1853^{+727}_{-528}$ & - & $2101^{+703}_{-501}$ & - \\
$\Delta {\rm BIC}$\tnote{e} & $-0.9$ & $9.4$ & $-17.6$ & $8.8$ \\
${\rm EW}_{\rm min, total}/\text{\AA}$\tnote{f} & $818^{+186}_{-159}$ & $446^{+123}_{-103}$ & $334^{+98}_{-70}$ & $227^{+49}_{-43}$ \\
$\log\left(M_{\rm BH, H\alpha}/M_\odot\right)$\tnote{g} & $7.1^{+0.4}_{-0.4}$ & - & $7.3^{+0.4}_{-0.4}$ & - \\
\hline
\end{tabular}
\begin{tablenotes}
\footnotesize
\item[a] Upper limits are based on the $3\sigma$ noise level.
\item[b] The system photometric redshift estimated by simply multiplying $z_{\rm photo}$ probability distribution function (PDF) of each component. We do not estimate $z_{\rm system}$ for CW-B2-4383 because the photometric redshift PDFs of each component differ significantly. However, note that we spectroscopically confirm that these components are at least at the same redshift.
\item[c] Based on the SED fitting with the redshift fixed to $z_{\rm H\alpha}$.
\item[d] Black hole masses estimated from the fitted blackbody luminosity assuming Eddington-limited accretion.
\item[e] BIC difference defined as one with the double Gaussian model minus one with the single Gaussian model.
\item[f] The lower-limit rest-frame H$\alpha$ EW estimated from the total line flux and the F444W photometry. For component \#1 of CW-B2-4383 and CW-A4-16093, we use the double-Gaussian results, while for component \#2 we use the single-Gaussian results.
\item[g] Black hole masses estimated from the fitted broad H$\alpha$ assuming single-epoch method.
\end{tablenotes}
\end{threeparttable}
\end{table*}

\subsection{CW-B5-15958}\label{ss:cw-b5-15958}

\begin{figure*}
 \begin{center}
  \includegraphics[width=17.8cm]{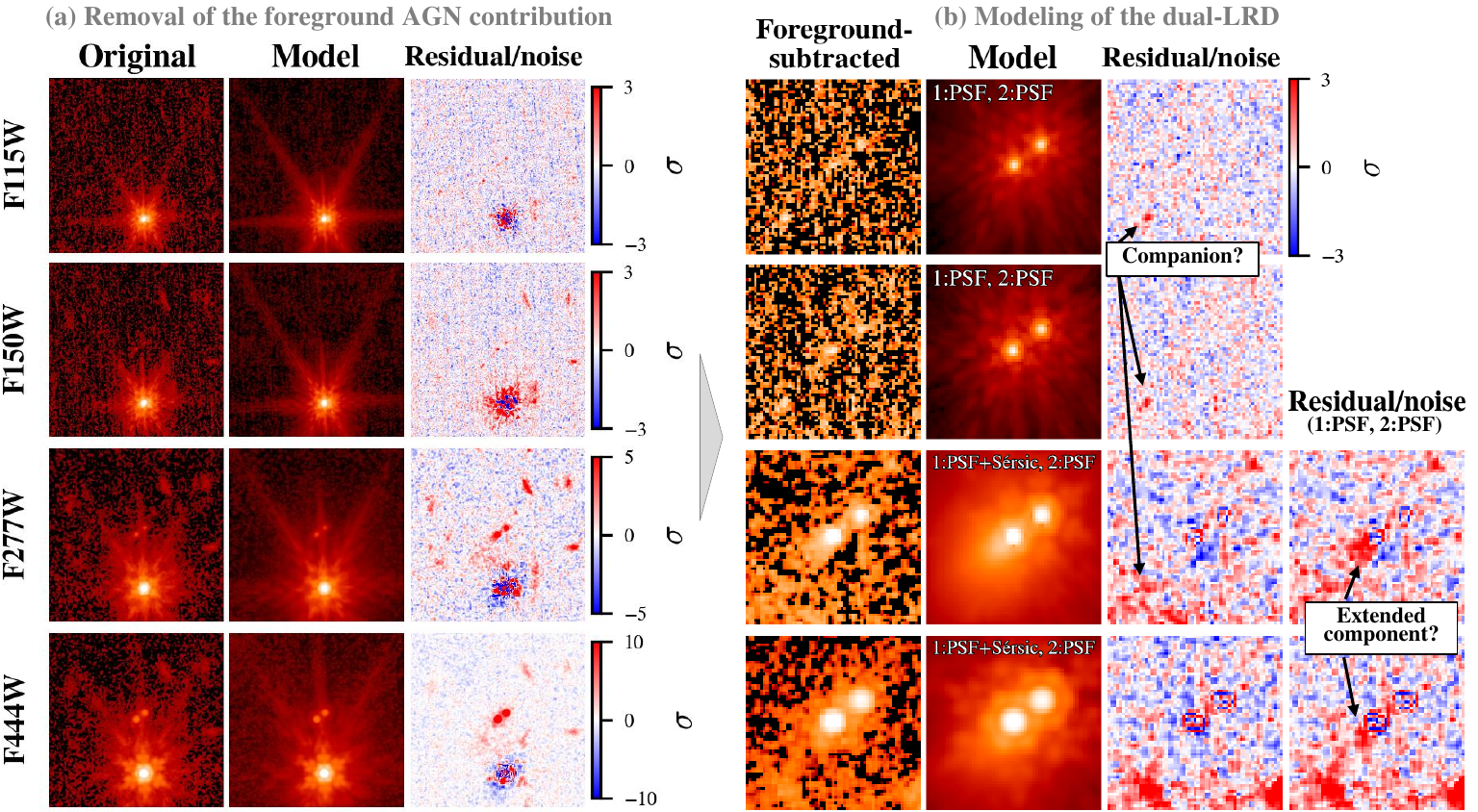} 
 \end{center}
\caption{The image-based modeling analysis on CW-B5-15958.
(a) Modeling of CID-643 to remove the foreground contribution by its AGN and host galaxy.
The left, middle, and right columns show the $6^{\prime\prime}\times6^{\prime\prime}$ original images, the best-fit model, and the residual (original - model) scaled by the noise map in each filter.
Note that, here, the residual image is Original - CID-643 (AGN and host component), and we do not subtract the CW-B5-15958 components.
(b) Modeling of CW-B5-15958 after subtracting CID-643 contribution.
Same for the panels (a), but focusing on CW-B5-15958 with the image size of $1\farcs8\times1\farcs8$.
In F277W and F444W, the best model is the composite model of double PSFs for each LRD and Sérsic for the extended component around component \#1.
For comparison, we also plot the scaled residual image with the PSF-PSF model (without the additional extended component). 
}
\label{fig:decom_b5_15958}
\end{figure*}

CW-B5-15958 is located 1\farcs8 away from an X-ray-detected AGN at $z_{\rm spec}=2.03$, CID-643, whose PSF spike overlaps with the emission from CW-B5-15958 as shown in the bottom-right panel of figure\,\ref{fig:3col_sed} (b) and figure\,\ref{fig:decom_b5_15958} (a).
Therefore, before performing image-based modeling of CW-B5-15958, we first subtracted the contribution of CID-643 from the original images.
In this step, we model CID-643 with a composite model of a PSF component representing the AGN and a Sérsic component for the host galaxy. 
For F277W and F444W, we also add two PSF components at the position of CW-B5-15958 to prevent its influence on the CID-643 modeling.
Figure\,\ref{fig:decom_b5_15958} (a) displays this subtraction analysis, where the residual images show the noise-scaled image after subtracting only the CID-643 components, i.e., CW-B5-15958 components are not subtracted even though they are simultaneously modeled in F277W and F444W, suggesting that the foreground contamination are effectively removed from the original images.

Next, we perform image-based modeling on the foreground subtracted images following the same procedure used for other objects (figure\,\ref{fig:decom_b5_15958} (b)).
In all filters, both components are better described by PSFs rather than Sérsic profiles.

The $m_{\rm F277W} - m_{\rm F444W}$ colors for components \#1 and \#2 are $1.9$ and $1.8$, respectively.
The optical slopes $\beta_{\rm opt}$, estimated from F444W and F277W, are $\beta_{\rm opt}=1.80\pm0.04$ and $1.57\pm0.10$ for components \#1 and \#2, respectively, similar to the slopes observed in typical LRDs \citep{Matthee2024, Greene2024, Kocevski2024}. 
Additionally, components \#1 and \#2 have $m_{\rm F115W} - m_{\rm F150W}$ colors of $0.3$ and $-0.1$, indicating a blue excess. 
These color differences suggest that components \#1 and \#2 are not gravitationally lensed images of the same object.
Both SEDs exhibit an LRD-like V-shape, well fitted with the LRD model.
The estimated photometric redshifts of components \#1 and \#2 are $z_{\rm photo}=6.30^{+0.59}_{-0.84}$ and $z_{\rm photo}=6.15^{+0.84}_{-1.27}$, respectively.
Despite the large uncertainty, these photometric redshifts are consistent with the two components residing at the same redshift.
We then combine their photometric redshift probability distribution functions (PDFs) and calculate the median and 68\% confidence interval of the combined distribution as $z_{\rm photo}=6.41^{+0.51}_{-0.70}$, which we assign as the photometric redshift of the system, $z_{\rm system}$.
The projected separation between the components \#1 and \#2 is $\theta=0\farcs37$ in F277W, corresponding to the physical separation of $d_{\rm sep} = 2.03^{+0.14}_{-0.08}\,{\rm kpc}$ at $z_{\rm system}$.

Interestingly, we find a $>3\sigma$ off-centered extended feature south of component \#1, as shown in the positive residuals after subtracting the fitted PSF-PSF model in F277W and F444W (the bottom-right panel of figure\,\ref{fig:decom_b5_15958} b). 
Since no similar structure is observed in the symmetric position relative to the PSF spike structure, it is unlikely that this extended component is a PSF diffraction feature that is not perfectly removed from the foreground subtraction.
To account for this extended feature in the image-based modeling, we fit component \#1 with a composite model of PSF and Sérsic, while component \#2 is fitted with a PSF.
This composite model results in significantly lower BIC values in both F277W and F444W compared to the PSF-PSF model.
The center of this extended component has a 0\farcs20 offset from component \#1, which corresponds to $1.03^{+0.10}_{-0.04}\,{\rm kpc}$ at $z_{\rm system}$.
Extended components around LRDs are found for other LRDs selected with the pixel-by-pixel selection method (Tanaka et al. in preparation).
Recently, \cite{Chen2024} reported that four out of eight spectroscopically confirmed and less-gravitationally-lensed LRDs in UNCOVER fields show off-centered extended components.
Unless coincidentally projected onto another galaxy, brown dwarfs are not expected to exhibit such extended features.
Therefore, this finding further supports that CW-B5-15958-\#1 is an LRD rather than a brown dwarf.
This extended component may represent an underlying host galaxy disturbed by interactions (e.g., tidal tail), a companion galaxy merging with the LRDs, or an ionization cone with nebular emission.

We also find another separated companion detected in F115W and F150W with the separation of 0\farcs76 (corresponding $4.18^{+0.28}_{-0.18}\,{\rm kpc}$ at $z_{\rm system}$) from component \#1.
This companion is undetected in F814W, and assuming it has a Lyman$\alpha$ break between F814W and F115W, it is consistent with being at the same redshift.
Therefore, this system is likely to consist of two LRDs, one with nearby extended components, and an additional companion galaxy located approximately $\sim4\,{\rm kpc}$ away (in projection) at the same redshift.
To spectroscopically confirm the dual LRD (i.e., obtaining spectroscopic redshifts) and explore the nature of the off-centered extended components and the companion further, spatially resolved deep spectroscopy with JWST/NIRSpec IFU is essential.

\subsection{CW-A6-19978}\label{ss:cw-a6-19978}
As shown in figure\,\ref{fig:3col_sed} (A), CW-A6-19978 has a fainter component \#2 located to the north of component \#1.
Using the F277W and F444W images, CW-A6-19978 is best modeled with a single PSF for each component based on a significantly lower BIC than other models.
The $m_{\rm F277W} - m_{\rm F444W}$ colors for components \#1 and \#2 are 1.7 and 1.8, respectively, and the optical slopes $\beta_{\rm opt}$ (where $F_\lambda\propto\lambda^\beta$) estimated from these two filters are $\beta_{\rm opt}=1.33\pm0.07$ and $1.57\pm0.28$ for components \#1 and \#2, respectively, similar to the optical slope observed in typical LRDs \citep{Matthee2024, Greene2024, Kocevski2024}.

With the F115W and F150W images, the best-fit model has a significantly lower BIC when using only one PSF for component \#1, i.e., no model for component \#2.
Therefore, we conclude that component \#2 is undetected in F115W and F150W. 
Component \#1 has the $m_{\rm F115W} - m_{\rm F150W}$ color of $-0.1$, indicating a blue excess (excess in $F_\lambda$ and flat slope in $F_\nu$).
SED fitting analysis with the LRD model results in the photometric redshift of component \#1 and \#2 as $z_{\rm photo}=6.36^{+0.68}_{-1.08}$ and $z_{\rm photo}=6.23^{+1.43}_{-1.43}$, resulting the photometric redshift of this system as $z_{\rm system} = 6.42^{+0.52}_{-0.63}$. 
When fitting component \#2, an upper limit on the photometry is used for the F115W and F150W bands.
The separation between the components \#1 and \#2 is $\theta=0\farcs21$ in F277W, the smallest among the three candidates in this study. 
The projected separation corresponds to $d_{\rm sep} = 1.15^{+0.08}_{-0.05}\,{\rm kpc}$ at $z_{\rm system}$.
In summary, both components are well-explained as unresolved components with a similar SED (figure\,\ref{fig:3col_sed}) to typical LRDs, with component \#2 detected only in F277W and F444W.

\subsection{CW-B2-4383}\label{ss:cw-b2-4383}
As shown in figure\,\ref{fig:3col_sed} (a), CW-B2-4383 has a fainter component \#2 located $0.\!\!^{\prime\prime}28$ away from component \#1.
A model with a PSF for component \#1 and a Sérsic for component \#2 has a significantly lower BIC than other model configurations for the F200W, F277W, F356W, and F444W images.
The best-fit Sérsic components have $r_e\in \left[0.\!\!^{\prime\prime}05, 0.\!\!^{\prime\prime}1\right]$ and $n\in \left[0.7, 5.0\right]$.
Although a Sérsic component is considered to be a more appropriate model, the results still suggest a very compact morphology.
However, the F277W image cutout (figure\,\ref{fig:3col_sed}) shows a slightly extended structure, which may be the result of its lower luminosity in F277W and potential contributions from extended components, as observed in CW-B5-15958 (section\,\ref{ss:cw-b5-15958}).
The $m_{\rm F277W} - m_{\rm F444W}$ colors for components \#1 and \#2 are 1.6 and 1.5, respectively.
The optical slope $\beta_{\rm opt}$ estimated from F444W and F277W are $\beta_{\rm opt}=1.09\pm0.15$ and $0.62\pm0.21$ for components \#1 and \#2, respectively, similar to the optical slope observed in typical LRDs \citep{Matthee2024, Greene2024, Kocevski2024}. 
A model composed of only one PSF for component \#1 has a significantly lower BIC than other models for the F115W and F150W images.
Thus, we conclude that component \#2 is undetected in F115W and F150W.
Component \#1 has $m_{\rm F115W} - m_{\rm F150W}$ color of $-0.2$, indicating a blue excess.

CW-B2-4383 is also covered by F770W.
In this band, the PSF-PSF model yields a significantly lower BIC compared to other configurations.
However, due to the limited spatial resolution of MIRI, the two components are almost entirely blended, as shown in figure\,\ref{fig:3col_sed}\,(b).
In addition, we find another potential companion detected in F277W, F356W, and F444W as indicated in figure\,\ref{fig:3col_sed}. 

\begin{figure*}
 \begin{center}
  \includegraphics[width=16cm]{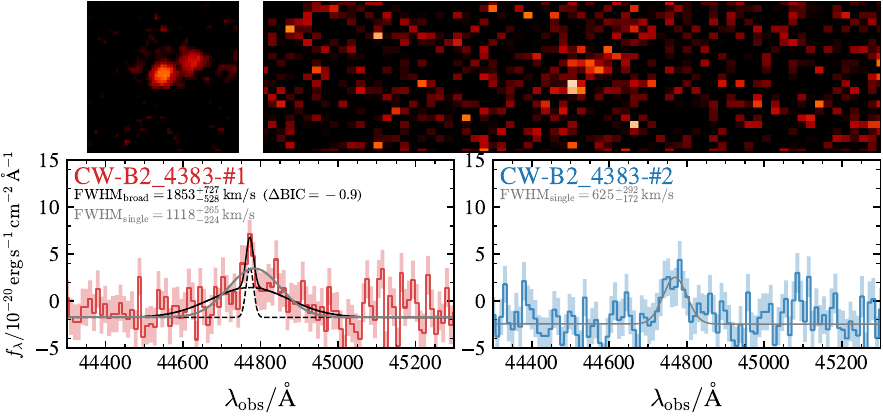} \end{center}
\caption{
F444W NIRCam image, COSMOS-3D 2D spectrum, and extracted 1D spectra of CW-B2-4383. 
In both the NIRCam image and the 2D spectrum, component \#1 is located at the center. 
In the 1D spectrum panels, the single- and double-Gaussian fitting results for component \#1 are shown with gray and black curves, respectively, while the single-Gaussian fit for component \#2 is shown with a gray curve. 
The corresponding FWHM values are indicated in the upper-left corner of each panel.
}
\label{fig:C3D_B2_4383}
\end{figure*}

The 2D spectrum (figure\,\ref{fig:C3D_B2_4383}) reveals two spatially separated lines detected at the same wavelength.
Given the very close separation, we extract the 1D spectra of each component through optimal extraction method.
The line-fitting results for each spectrum are summarized in table\,\ref{tab:dual_LRD_spec}. 
In both components, the emission line is detected at the same wavelength of $4.478\,{\rm \mu m}$.
The detection significance of the line, evaluated as the $S/N$ based on the estimated total line luminosity and its $1\sigma$ uncertainty, is $S/N=4.5$ and $3.6$ for component \#1 and \#2, respectively.

For component \#1, the double-Gaussian fitting yields a lower BIC than the single-Gaussian model with the broad component FWHM of $1853^{+727}_{-528}\,{\rm km\,s^{-1}}$.
However, the BIC difference $\Delta {\rm BIC}$, defined as BIC for the double-Gaussian minus the BIC for the single-Gaussian, is only $\Delta{\rm BIC}=-0.9$, which does not provide strong evidence to select the double-Gaussian model, possibly due to the low $S/N$.
Nevertheless, even in the single-Gaussian fitting, the line FWHM is relatively broad with ${\rm FWHM_{single}}= 1118^{+265}_{-224}\,{\rm km\,s^{-1}}$), suggesting the presence of a broad-line feature in component \#1.
For component \#2, the single-Gaussian model yields a lower BIC than the double-Gaussian model, with a moderately broad line width of ${\rm FWHM_{single}}=625^{+292}_{-172}\,{\rm km\,s^{-1}}$.
However, given the low $S/N$ and the potential contamination from component \#1, it is difficult to robustly determine whether component \#2 truly hosts a broad-line feature.

Due to single-line detections, we cannot completely rule out the possibility that these lines are intrinsically different lines from different redshifts that coincidentally appear at the same wavelength.
However, the likelihood of a chance projection of two LRDs at different specific redshifts that would produce different emission lines appearing at the same observed wavelength, with a projected separation of $0.\!\!^{\prime\prime}2$, is extremely low (see also Section\,\ref{s:discussion}).
Thus, it is highly probable that this system is a genuine pair at the same redshift.

The rest-frame EW estimated from the total line flux and the F444W photometry are $818_{-159}^{+186}\,\text{\AA}$ and $446^{+123}_{-103}\,\text{\AA}$.
While we use the single-Gaussian fitting result to estimate the EW of component \#2, the limited $S/N$ may prevent the detection of a broad component.
Therefore, the derived EW for component \#2 should be regarded as a lower limit.
Another possible way to constrain the EW is to estimate an upper limit on the broad-line flux from mock analyses and then derive a corresponding upper limit on the EW. 
However, our mock analysis shows that it is challenging to place a meaningful upper limit on the broad-line flux because the FWHM is also unknown (see appendix\,\ref{ap:broad_line}).
Given these strong line features and tentative detection of a broad component, potential lines are C{\sc iv} ($z=27.90$), Mg{\sc ii} ($z=14.99$), H$\beta$ ($z=8.210$), H$\alpha$ ($z=5.822$), Pa$\beta$ ($z=2.495$), or Pa$\alpha$ ($z=1.388$).
Among them, C{\sc iv}, Mg{\sc ii}, Pa$\beta$, and Pa$\alpha$ can be rejected because the inferred redshift is inconsistent with the photometric redshift of both component.
If it is H$\beta$, [O{\sc iii}]$\lambda\lambda4959,5007$ doublets, which can be typically seen in LRD spectrum \citep[e.g.,][]{Greene2024}, should appear in $\lambda_{\rm obs}=4.567\,{\rm \mu m}$ and $4.612\,{\rm \mu m}$.
However, we cannot see any features even in component \#1.
Thus, we conclude that this line is likely H$\alpha$ at $z_{\rm H\alpha}
=5.822$.
Additionally, the $z_{\rm H\alpha}$ matches the photometric redshift of component \#1 ($z_{\rm photo}=5.66^{+0.42}_{-0.35}$).
At $z_{\rm H\alpha}$, the projected separation between \#1 and \#2 is $d_{\rm sep} = 1.64\,{\rm kpc}$.

Due to the inclusion of F770W photometry and the fixed $z$, CW-B2-4383 exhibits smaller uncertainties on the inferred parameters from SED fitting.
The H$\alpha$ EWs inferred from the SED fitting are $\left(6.1_{-3.0}^{+4.0}\right)\times10^2\,\text{\AA}$ and $\left(7.5_{-4.9}^{+7.5}\right)\times10^2\,\text{\AA}$ for components \#1 and \#2, respectively. 
Although only the broad-band photometry is available and the EW is degenerate with $T_{\rm eff}$ and the blackbody luminosity, the inferred values are consistent with the lower-limit EW derived from the spectroscopic data analysis discussed above. 
To better constrain the intrinsic SED and break these degeneracies, deeper spectroscopic observations are required.

CW-B2-4383 is also covered by ALMA in Band 3 ($89.5\,\mathchar`-\,105.4\,{\rm GHz}$, Project ID: 2016.1.00171.S) with a continuum sensitivity of $0.0395\,{\rm mJy/beam}$. 
We do not find any detections in the continuum or signs of emission lines in the cubes.
This non-detection is consistent with previous studies that have reported the absence of LRD detections in ALMA observations \citep{Labbe2023b, Akins2024}.

\subsection{CW-A4-16093}\label{ss:cw-a4-16093}
CW-A4-16093 has the largest projected separation in our sample, $1.\!\!^{\prime\prime}23$ (figure\,\ref{fig:3col_sed} a). 
Component \#1 was also selected in \cite{Akins2024}, whereas component \#2 was not. 
This difference is likely due to their colors.
Component \#1 has $m_{\rm F277W} - m_{\rm F444W}=1.9$, while component \#2 has $m_{\rm F277W} - m_{\rm F444W}=1.4$, which lies close to the selection threshold.

From F115W to F770W, modeling both components with a PSF yields significantly lower BIC values than alternative model configurations, indicating that both components are compact object consistent with point source.
The optical slopes, $\beta_{\rm opt}$, estimated from F277W and F444W are $\beta_{\rm opt}=1.77\pm0.15$ and $0.89\pm0.16$ 
for components \#1 and \#2, respectively, comparable to those observed in typical LRDs \citep{Matthee2024,Greene2024,Kocevski2024}. 
The $m_{\rm F115W} - m_{\rm F150W}$ colors of components \#1 and \#2 are each $-0.2$, respectively, indicating a blue excess.

We identify another bright companion detected from F115W to F444W, as shown in figure\,\ref{fig:3col_sed}. 
However, this object is also detected in the Subaru HSC $g$-band image, making it unlikely to be at the same redshift as the LRDs.
In addition, particularly in the F200W and F356W COSMOS-3D imaging relatively deeper than the COSMOS-Web images, we detect another faint, blob-like companion. 
However, due to its faintness, it is difficult to estimate reliable physical properties such as a photometric redshift.

\begin{figure*}
 \begin{center}
  \includegraphics[width=16cm]{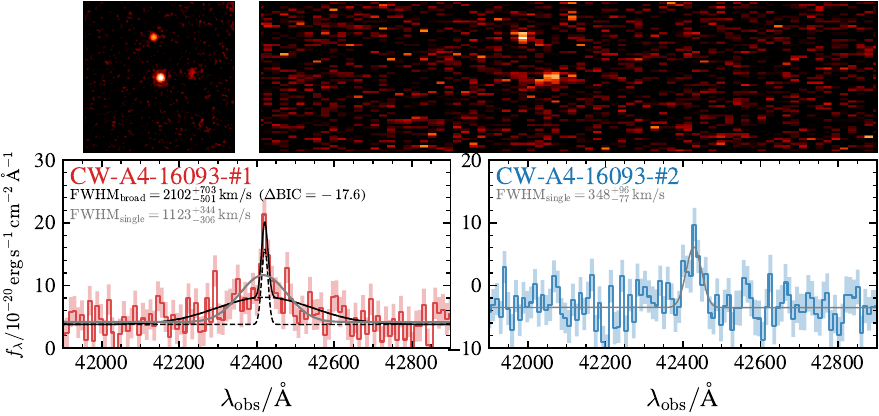} \end{center}
\caption{
Same as figure\,\ref{fig:C3D_B2_4383}, but for CW-A4-16093.
}
\label{fig:C3D_A4_16903}
\end{figure*}

Thanks to the sufficiently large separation, the 2D spectrum (figure\,\ref{fig:C3D_A4_16903}) reveals two emission lines that are more clearly spatially separated than in the case of CW-B2-4383, detected at the same wavelength.
The line-fitting results for each spectrum are summarized in table\,\ref{tab:dual_LRD_spec}. 
In both components, the emission line is detected at the same wavelength of $4.242\,{\rm \mu m}$.
The detection significance of the line is $S/N=3.7$ and $3.9$ for component \#1 and \#2, respectively.

For component \#1, the double-Gaussian fitting results in  the broad component FWHM of ${\rm FWHM_{broad}}=2101^{+703}_{-501}\,{\rm km\,s^{-1}}$ with significantly lower BIC of $\Delta{\rm BIC}=-17.6$ than the single Gaussian model with ${\rm FWHM_{single}}=1123^{+344}_{-306}\,{\rm km\,s^{-1}}$.
For component \#2, the single-Gaussian model yields a lower BIC than the double-Gaussian model, with ${\rm FWHM_{\rm single}}=348^{+96}_{-77}\,{\rm km\,s^{-1}}$.
Based on this result alone, component \#2 may appear to lack a broad emission line and thus may not qualify as an LRD. 
However, as discussed in Appendix\,\ref{ap:broad_line}, the possibility that a broad component remains undetected due to the low $S/N$ cannot be ruled out.

The single-line detections do not completely rule out the possibility that the two different lines at different redshifts coincidentally appear at the same observed wavelength.
However, as in the case of CW-B2-4383, the probability of such a chance projection within a projected separation of $0.\!\!^{\prime\prime}2$ is extremely low (see also section\,\ref{s:discussion}). 
We therefore conclude that this system is very likely a genuine pair at the same redshift.

The lower-limit rest-frame EWs estimated from the line flux and the F444W photometry are $492\,\text{\AA}$ and $251\,\text{\AA}$ for components \#1 and \#2, respectively. 
Given the spectroscopic constraints, the photometric redshift of the system ($z_{\rm system}=5.61^{+0.30}_{-0.31}$), and the lack of [O{\sc iii}]-like features, we conclude that the detected line is most likely H$\alpha$ at $z_{\rm H\alpha}=5.464$. 
At this redshift, the projected separation between \#1 and \#2 corresponds to $d_{\rm sep} = 7.36\,{\rm kpc}$.

Due to the inclusion of F770W photometry and the fixed $z$, CW-A4-16093 exhibits smaller uncertainties on the inferred parameters from SED fitting.
The H$\alpha$ EWs inferred from the SED fitting are $\left(4.8_{-2.7}^{+3.4}\right)\times10^2\,\text{\AA}$ and $\left(9.3_{-4.7}^{+6.3}\right)\times10^2\,\text{\AA}$ for components \#1 and \#2, respectively.
These results are consistent with the lower-limit EW derived from the spectroscopic data analysis discussed above. 

\section{Excess clustering on kpc-scales}\label{s:discussion}

\begin{figure}
 \begin{center}
  \includegraphics[width=8.5cm]{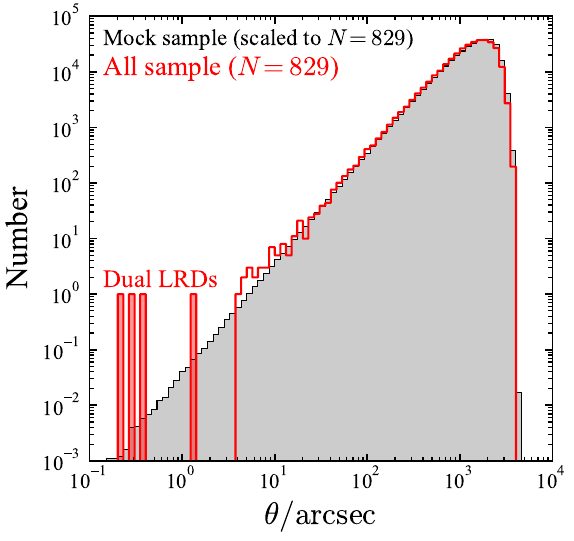}
 \end{center}
\caption{
Distribution of projected separation ($\theta$) for all and mock samples.
Red histogram indicates the $\theta$ distribution for an LRD sample selected with pixel-by-pixel color selection ($N=829$). 
Gray histogram indicates the mock LRD sample (scaled to $N=829$) generated with a random spatial distribution in the COSMOS-Web-like field.
The red filled rectangle corresponds to our dual LRD sample presented in this paper, which indicates the challenge to explain our dual LRDs as a pair of two physically unrelated objects (LRDs or BDs) at different redshifts.
}
\label{fig:separations1}
\end{figure}

In this section, we compare our dual LRD sample with the full LRD sample selected using the pixel-by-pixel color selection method, as well as with samples from previous studies, to investigate whether the discovery of dual LRDs implies enhanced clustering of LRDs on kpc scales. 
The sample selected using the pixel-by-pixel color selection method from the COSMOS-Web field includes dusty galaxies and brown dwarfs that exhibit very red colors similar to LRDs. 
To mitigate such contamination, we construct a cleaned sample by selecting objects that are classified as LRD-like based on the dimensionality-reduction analysis, are better fitted by a PSF or PSF+Sérsic model rather than a pure Sérsic model, and are not identified as brown dwarfs from their SED fitting results.
We also restrict the sample to $S/N > 10$ in F444W, because mock analyses show that the completeness drops sharply at $S/N \lesssim 10$. 
With these selections, the final sample size becomes $N = 829$.
Due to the inclusion of previously overlooked LRDs, such as LRDs with close companions and fainter LRDs, this sample is approximately twice as large as that of \citep[][$N=434$]{Akins2024}, who conducted an aperture-based photometric search for LRDs in COSMOS-Web using the same color threshold of $m_{\rm F277W}-m_{\rm F444W}>1.5$.
A detailed analysis of this sample will be presented in paper\,II (Tanaka et al., in preparation).
Using this full LRD sample, We first assess and rule out the possibility of chance projection, and then examine the evidence for excess clustering.

\subsection{Probability of a chance projection}\label{ss:coincident1}
For all possible LRD pairs from the full LRD sample, we measure the distribution of their projected separations on the sky. 
We then generate 10,000 mock LRD samples with the same sample size as the full LRD sample ($N=829$) randomly distributed within the COSMOS-Web field, which is simply modeled as a square region with an area of $0.54\,{\rm deg^2}$. 
For each mock sample, we compute the separation distribution in the same manner.

The observed and mock separation distributions are broadly consistent (figure\,\ref{fig:separations1}), indicating that the global distribution of LRDs can be explained by a random distribution. 
The Monte Carlo analysis using 10,000 mock samples shows that the probability of finding four pairs with $\theta\lesssim1.\!\!^{\prime\prime}2$ is $\sim7\times10^{-5}$ (corresponding to $3.8\sigma$). 
Similarly, the probability of finding three pairs with $\theta\lesssim0.\!\!^{\prime\prime}4$ is $\sim4\times10^{-6}$ (corresponding to $4.5\sigma$).

In section\,\ref{s:dual}, we have argued that contamination from brown dwarfs is highly unlikely in our dual LRD sample from the SED and the presence of offset components or companion galaxy candidates. 
Nevertheless, even if we conservatively consider the possibility of BD contamination, the conclusion regarding chance projection remains unchanged.
In our parent sample, 77 objects are excluded as BD candidates based on their SEDs, and the overall BD contamination fraction is generally estimated to be $\sim10\%$ \citep[e.g.,][]{Akins2024, Greene2024}. 
Adopting a more conservative contamination fraction of $50\%$, we artificially increase the mock sample size to $N=1,660$ and repeat the Monte Carlo analysis. 
Even under this assumption, the probability of finding four pairs with $\theta\lesssim1.\!\!^{\prime\prime}2$ remains $\sim2\times10^{-4}$ (corresponding to $3.5\sigma$), and the probability of finding three pairs with $\theta\lesssim0.\!\!^{\prime\prime}4$ remains $\sim7\times10^{-6}$ (corresponding to $4.3\sigma$).
Therefore, we conclude that the dual LRD candidates are unlikely to be explained by chance projection.

In addition, another interesting point here is that, even when using the full sample, we do not find any other pairs with $\theta=0\farcs4\,\mathchar`-\,3\farcs8$ (corresponding to $\sim2\,\mathchar`-\,20\,{\rm kpc}$ at $z\sim6$) than CW-A4-16093.
For $\theta\gtrsim0\farcs5$, the influence of a companion LRD on compactness measurements should be negligible (also see Paper II), meaning LRDs with a $\theta\sim1^{\prime\prime}$ companion should be detectable with typical selection methods.
However, even when using the sample of photometrically selected LRDs from COSMOS-Web identified in a conventional manner by \cite{Akins2024}, no objects with similarly small separations are found.
This absence may suggest that the triggering of LRD activity or LRD formation through interactions works only on very small scales ($\sim0\farcs4$), corresponding to $1.9\,\mathchar`-\,2.5\,{\rm kpc}$ at $z=5\,\mathchar`-\,8$.
Considering that LRDs may have low stellar masses such as $M_*\lesssim10^{9}M_\odot$ \citep{Chen2024, Zhang2025_host} or $M_*\sim5\times10^7 M_\odot$ \citep{matthee2024_env}, the spatial scale at which AGN activity is triggered by interactions in LRDs may be smaller than that for typical low-$z$ AGNs.
Expanding survey fields and performing spectroscopic identification of LRDs are crucial for constructing a larger sample and understanding their clustering and interaction-driven triggering.

\subsection{Excess clustering on kpc scales}\label{ss:dual_number2}
\begin{figure}
 \begin{center}
  \includegraphics[width=8.5cm]{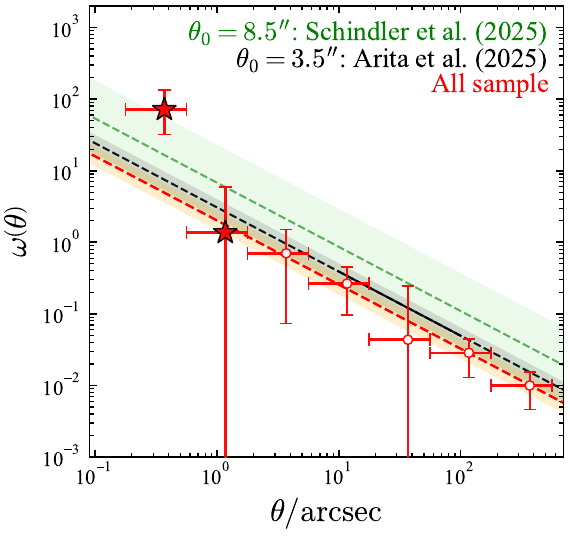} 
 \end{center}
\caption{Angular auto-correlation function of LRDs integrated at $z=5\,\mathchar`-\,8$.
The red plots indicate the angular auto-correlation function derived from the full sample, with the star-shaped symbols corresponding to the dual LRDs reported in this paper.
Note that many LRDs in the full sample and the two of the four dual LRD candidates are without spectroscopic confirmation.
Black/green lines and gray/green shaded regions indicate the angular auto-correlation function estimated based on the results by \cite{Arita2024} and \cite{Schindler2024}.
}
\label{fig:acf}
\end{figure}

\subsubsection{Angular auto-correlation function}\label{sss:comp_Akins24}
The presence of such closely separated systems, which cannot be explained by chance projection, suggests that the dual LRD candidates reside at the same redshift as physically associated, merging systems, and that LRDs exhibit enhanced clustering on small spatial scales.
To further quantify this, we utilize the full LRD sample to constrain the angular auto-correlation function (ACF).

We make 82,900 (100 times the sample size) random points over the COSMOS-Web region.
Then, following the method of \cite{Arita2024}, we evaluate the angular ACF using the estimator from \cite{Landy1993};
 \begin{equation}
     \omega\left(\theta\right) = \frac{LL\left(\theta\right) - 2LR\left(\theta\right) + RR\left(\theta\right)}{RR\left(\theta\right)},
 \end{equation}
where $LL\left(\theta\right)$, $LR\left(\theta\right)$, and $RR\left(\theta\right)$ represent the normalized pair counts of LRDs - LRDs, LRDs - random points, and random points - random points within the $\theta$ range.
For simplicity, we ignore the integral constraints.
The errors are estimated assuming Poisson statistics.

The resulting ACF, shown in blue plots in figure\,\ref{fig:acf}, generally aligns with the prediction from \cite{Arita2024} described later.
Following \cite{Arita2024}, we fit the ACF between $\theta=2\,\mathchar`-\,600^{\prime\prime}$ with 
\begin{equation}
    \omega\left(\theta\right) = \left(\frac{\theta}{\theta_0}\right)^{-\beta}, \label{eq:omega}
\end{equation}
where $\beta$ is fixed to $\beta=0.90$ for comparing with different studies (section\,\ref{sss:comp_Mhalo}).
The fitted power law with $\theta_0=2\farcs2^{+0\farcs9}_{-0\farcs6}$ is shown in the red dashed line in figure\,\ref{fig:acf}, and over 68\% confidence intervals, the angular ACF has an excess at very small scale of $\theta\lesssim1^{\prime\prime}$ from the expectation of fitted power law.
Especially, in the smallest angular bin at $\theta \sim 0.\!\!^{\prime\prime}2$, the measured ACF is $47_{-27}^{+40}$ times higher than that expected from the fitted power-law model.
Note that we use bins of $\Delta\log\left(\theta/{\rm arcsec}\right)=0.5$ for the above results.
We also confirmed that the small-scale excess of a factor of several tens is robust against the choice of bin width.

\subsubsection{Comparison with ACFs from previous studies}\label{sss:comp_Mhalo}
Some studies have conducted clustering analyses of LRDs and estimated their halo mass ($M_{\rm halo}$, e.g., \citealt{Pizzati2024}). 
\cite{Arita2024} conducted a clustering analysis of low-luminosity JWST-found broad-line (${\rm FWHM} \gtrsim 1000\,{\rm km/s}$) AGNs at $z\sim5\,\mathchar`-\,6$, including LRDs, and estimate the halo mass of $\log (M_{\rm halo}/h^{-1}M_\odot) = 11.46_{-0.25}^{+0.20}$.
\cite{Lin2026_DMhalo} estimated the halo mass of JWST-discovered broad H$\alpha$ AGNs at $z=5\,\mathchar`-\,6$ to be $\log\left(M_{\rm halo}/M_\odot\right)=11.04^{+0.34}_{-0.32}$ by comparing their clustering measurements with predictions from UniverseMachine simulation \citep{Behroozi2019} and to be $\log\left(M_{\rm halo}/M_\odot\right)=11.76^{+0.26}_{-0.38}$ directly from the bias parameter obtained in the clustering analysis.
These results are lower than the halo mass of high-$z$ quasars \citep[$\log\left(M_{\rm halo}/h^{-1}M_\odot\right)\sim12\,\mathchar`-\,13$, e.g.,][]{Shimasaku2019, Arita2023, Cordova2024, Eilers2024}.
Note that the halo mass of LRDs is still in discussion, and \cite{Schindler2024} reported a minimum halo mass of $\log\left(M_{\rm halo}/h^{-1}M_\odot\right)=12.02^{+0.82}_{-1.00}$, similar to high-$z$ quasars.
However, we caution that this high $M_{\rm halo}$ result is based on a single LRD at $z\sim7.3$ embedded in an overdensity of eight galaxies.

The previous studies have measured spatial cross correlation functions (CCFs) between galaxies and JWST AGNs, whereas we measure the angular ACF for LRDs.
Therefore, the previously reported correlation lengths are for CCFs and cannot be directly compared with our measurements.
Instead, we predict the angular ACF from the results of the previous studies, considering that the conversion between the correlation length of the spatial ACF and halo mass, which depends on redshift.
We assume that LRDs reside in halos with similar halo masses across different redshifts, and convert the previously reported halo-mass measurements into the corresponding angular ACFs at $z=6$.
We convert the previously reported halo masses into the correlation length of the spatial ACF, $r_{\rm 0, AGN\,\mathchar`-\,AGN}$, assuming $z=6$ and following the relation by \cite{Arita2023}.
We then apply Limber's equation \citep{Limber1953}, which transforms the spatial ACF into the angular ACF, assuming the form of equation\,(\ref{eq:omega}) with $\beta=0.90$ \citep{Arita2024} and the redshift distribution of LRDs at $z=5\,\mathchar`-\,8$ \citep{Akins2024}.

When assuming $\log (M_{\rm halo}/h^{-1}M_\odot) = 11.46_{-0.25}^{+0.20}$ as measured by \cite{Arita2024}, we obtain $\theta_0=3\farcs5^{+1\farcs2}_{-1\farcs1}$.
This expectation (black line in figure\,\ref{fig:acf}) is consistent with the measured angular ACF at large separations of $\theta\sim10\,\mathchar`-\,300^{\prime\prime}$. 
Even when adopting this expectation as a reference, we still find a significant enhancement on small scales of $\theta<1^{\prime\prime}$ at the $68\%$ confidence level, with $\omega(0.\!\!^{\prime\prime}2)$ showing an excess of $23_{-11}^{+22}$ times relative to the expectation.
Note that the correlation function presented by \cite{Arita2024} is derived from JWST AGNs, including not only LRDs but also other types of AGNs. 
Therefore, the comparison may not be strictly equivalent if photometrically selected LRDs and spectroscopically found broad line AGNs occupy halos of different typical masses. 

If we assume a large halo mass of $\log\left(M_{\rm halo}/h^{-1}M_\odot\right) = 11.76^{+0.26}_{-0.38}$ and $\log\left(M_{\rm halo}/h^{-1}M_\odot\right) = 12.02^{+0.82}_{-1.00}$ estimated from \cite{Lin2026_DMhalo} and \cite{Schindler2024}, the correlation length is $\theta_0 = 5\farcs6^{+2\farcs9}_{-2\farcs5}$ and $8\farcs5_{-6\farcs4}^{+24\farcs7}$, respectively.
Even in these large halo mass cases, $\omega(0\farcs2)$ excess still exists (green line in figure\,\ref{fig:acf}).
However, the ACF at larger angular scales shows a clear offset from the model prediction, suggesting that such a large halo mass is unlikely to be representative of our sample.

In summary, our ACF measurements are broadly consistent, on large scales, with previous clustering analyses that report typical halo masses of $\log \left(M_{\rm halo}/M_\odot\right)\sim11\,\mathchar`-\,11.5$. 
However, on small scales of $\lesssim1^{\prime\prime}$, we find an excess clustering of a factor of $\sim20\,\mathchar`-\,50$ relative to the extrapolation from larger separations ($\theta\sim10\,\mathchar`-\,600^{\prime\prime}$).

If the ACF excess is special for LRDs, one possible interpretation is that LRDs are AGNs.
Such an excess in the ACF at small $\theta$ is qualitatively consistent with AGNs at lower redshifts \citep[e.g.,][]{Hennawi2006, Eftekharzadeh2017, Bhowmick2019, Shen2023, Comerford2024}, which may indicate that LRDs are AGNs and that close interactions trigger their activity.
However, normal galaxies also have excess ACFs on small scales.
Therefore, to compare with LRDs, we need to constrain the ACF of not only LRDs but also normal galaxies from a large-scale to kpc-scale in future observations covering large areas with JWST/NIRCam WFSS, such as COSMOS-3D and NEXUS \citep{Shen2024_NEXUS, Zhuang2024_NEXUS}.

Finally, all analyses here are based on photometrically selected dual LRD candidates and LRD sample, many of which lack spectroscopic confirmation.
Further spectroscopic identification is crucial to verify the arguments.
The sample size of candidates is still, at most, only three, and it is needed to create a more statistically significant sample from a large NIRCam imaging data \citep[e.g.,][]{morishita2024_beacon}.

\begin{figure*}
 \begin{center}
  \includegraphics[width=16cm]{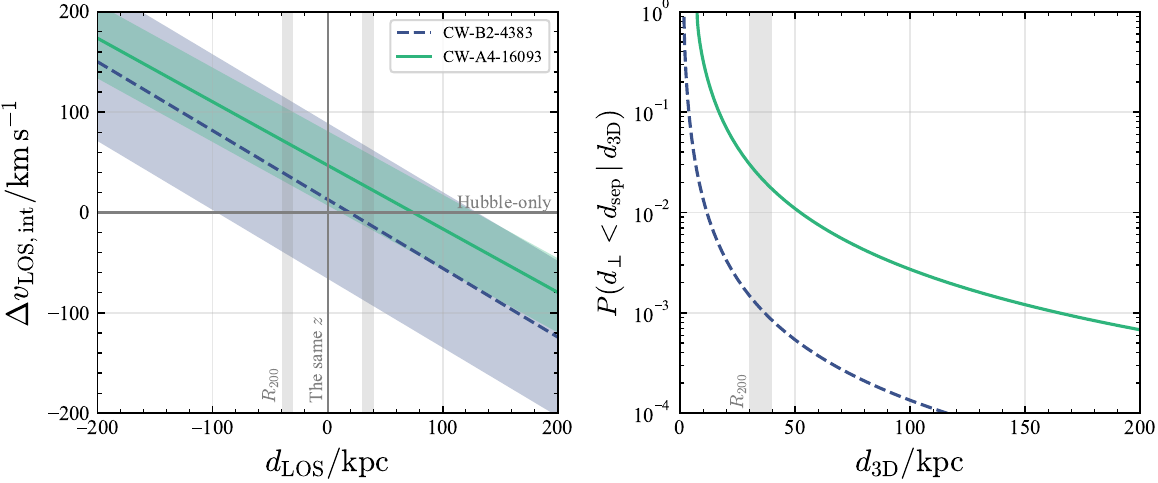}
  \end{center}
    \caption{
    (Left) Degeneracy between the intrinsic line-of-sight velocity offset, $\Delta v_{\rm LOS,int}$, and the line-of-sight separation, $d_{\rm LOS}$, inferred from the observed line-centroid offsets for CW-B2-4383 (purple, dashed) and CW-A4-16093 (green, solid).
    The shaded regions indicate the uncertainties corresponding to the uncertainties in the central wavelength offsets.
    The cases assuming the same cosmological redshift ($d_{\rm LOS}$) and a pure Hubble-flow interpretation ($\Delta_{\rm LOS, int}$) are indicated by the gray solid lines.
    The gray shaded region corresponds to the rough $R_{200}$ range of the $z\sim6$ LRD host halos.
    (Right) Probability that the projected separation $d_\perp$ becomes smaller than the observed projected separation $d_{\rm sep}$, as a function of the true three-dimensional separation $d_{\rm 3D}$, for CW-B2-4383 (purple, dashed) and CW-A4-16093 (green, solid).
    The observed systems are likely to have $d_{\rm 3D}$ less than $\sim R_{200}$ and intrinsic line-of-sight velocity offsets of less than $\sim 10^2\,{\rm km\,s^{-1}}$.
    }
\label{fig:separation_redshift}
\end{figure*}

\subsection{Non-detection of dual LRDs in other JWST fields}\label{ss:other_fields}
We also apply the pixel-by-pixel color selection method to NIRCam imaging data from the CEERS (\href{https://www.stsci.edu/jwst/science-execution/program-information?id=1345}{ERS\,1345}, PI: S. L. Finkelstein, \citealt{Finkelstein2025_ceers}), JADES (\citealt{Eisenstein2026_JADES}), and PRIMER-UDS (\href{https://www.stsci.edu/jwst/science-execution/program-information?id=1837}{GO\,1837}, PI: J. S. Dunlop) programs.
For CEERS and JADES, we use the NIRCam imaging data from CEERS Data Release version 1.0 \citep{Bagley2023, Finkelstein2025_ceers} and JADES Data Release 1 and 2 for GOODS-North and GOODS-South, respectively \citep{Rieke2023_JADES, Eisenstein2025_JADES}.
For PRIMER-UDS, we use imaging data processed in the same way as the COSMOS data adopted in this study (see section\,\ref{s:data_method}).

After applying the pixel-by-pixel color selection to these field data and performing visual inspection, we find no dual LRD systems (also see appendix\,\ref{ap:others} for a dubious system that we conclude is not a dual LRD).
The area covered by COSMOS-Web ($0.54\,{\rm deg^2}$ corresponding to $1.9\times10^3\,{\rm arcmin^2}$) is approximately $\sim20$, $\sim11$, and $\sim 8.3$ times larger than CEERS ($\sim100\,{\rm arcmin^2}$), JADES ($175\,{\rm arcmin^2}$), and PRIMER-UDS ($234\,{\rm arcmin^2}$), respectively.
The combined area of the CEERS, JADES, and PRIMER-UDS is still $\sim3.7$ times smaller than that of the COSMOS-Web.
Therefore, given that four dual LRDs are identified in COSMOS-Web, the absence of dual LRDs with at least comparable flux in the other fields is broadly consistent with the difference in survey area.
However, these surveys are deeper than COSMOS-Web and therefore probe intrinsically fainter LRDs. 
We plan to revisit this implication with consideration of the depth differences through comparing the luminosity functions of LRDs identified in each field in a forthcoming paper.

\section{Dual LRD as a BH merger precursor}\label{s:discussion2}

In this section, we discuss whether the discovered dual LRDs will eventually merge and, if so, what range of black hole masses will merge.

\subsection{Ruling out fly-by encounters}
First, we discuss the dynamical state of the spectroscopically confirmed dual LRD systems. 
For CW-B2-4383 and CW-A4-16093, the physical projected separations, $d_{\rm sep}$, at $z_{\rm H\alpha}$ are $1.64\,{\rm kpc}$ and $7.36\,{\rm kpc}$, respectively.
These projected separations are much smaller than the virial radius of their host dark-matter halos, which we characterize by $R_{200}$, the radius within which the mean enclosed density is 200 times the critical density at each redshift, $\rho_{\rm crit}\left(z\right)$:
\begin{align}
R_{200} &= \left( \frac{3 M_{\rm 200}}{800 \pi \rho_{\rm crit}(z)} \right)^{1/3} \notag,
\end{align}
where $M_{200}$ is the enclosed mass within $R_{200}$, and we assume $M_{\rm halo}\simeq M_{200}$.
For $M_{\rm halo}=10^{11\,\mathchar`-\,11.5}M_\odot$ at $z\sim6$ \citep{Arita2024, Lin2026_DMhalo}, $R_{200}$ takes $\simeq 30\,\mathchar`-\,40\,{\rm kpc}$.
Therefore, the observed projected separations of our dual LRD sample ($1\,\mathchar`-\,8\,{\rm kpc}$) correspond to only $\sim 5\,\mathchar`-\,10\%$ of $R_{200}$, indicating that the two systems reside well within the same halo.

From the central wavelengths of the fitted emission lines, $\mu_1$ and $\mu_2$, and assuming that the two LRDs are located at the same redshift, we estimate their line-of-sight velocity offset as
\begin{equation}
\Delta v_{\rm LOS}
= c\,\frac{\mu_1-\mu_2}{(\mu_1+\mu_2)/2}.
\end{equation}
We obtain $\Delta v_{\rm LOS} = 13^{+76}_{-79}\,{\rm km\,s^{-1}}$ and $47^{+34}_{-40}\,{\rm km\,s^{-1}}$ for CW-B2-4383 and CW-A4-16093, respectively.

However, the observed wavelength difference cannot be uniquely decomposed into an intrinsic line-of-sight velocity offset and a redshift difference (i.e., distance difference) since they are completely degenerated.
If part of the wavelength offset is attributed to the Hubble flow, the line-of-sight intrinsic velocity difference, $\Delta v_{\rm LOS, int}$, is given by
\begin{equation}
\Delta v_{\rm LOS, int} = \Delta v_{\rm LOS} - H\left(z\right)\,d_{\rm LOS},
\end{equation}
where $H\left(z\right)$ is the Hubble constant at each redshift, and $d_{\rm LOS}$ is the proper line-of-sight separation corresponding to the Hubble-flow contribution. 
In this case, the three-dimensional separation between two LRDs, $d_{\rm 3D}$, is
\begin{equation}
d_{\rm 3D} = \sqrt{d_{\rm LOS}^2 + d_{\rm sep}^2\,}\,.
\end{equation}

For example, if all of the observed wavelength offsets are from the Hubble flow, the inferred line-of-sight separations are $d_{\rm LOS}=19^{+111}_{-115}\,{\rm kpc}$ and $74^{+54}_{-63}\,{\rm kpc}$ for CW-B2-4383 and CW-A4-16093, respectively, as shown in figure\,\ref{fig:separation_redshift} (left). 
These values are approximately an order of magnitude larger than $d_{\rm sep}$, and are roughly comparable to the halo virial radius discussed above.
In even more extreme cases, the separations could formally exceed the virial radius. 

Nevertheless, since the viewing direction of a pair is random, the geometric probability of observing a projected separation much smaller than the true three-dimensional separation becomes small.
We define the viewing angle $\theta$ as the angle between the line connecting the two LRDs and the line of sight.
For a pair with true separation $d_{\rm 3D}$, the viewing angle $\theta$ required to observe a projected separation $d_{\perp}<d_{\rm sep}$ is
\begin{equation}
\theta \leq \arcsin\left(\frac{d_{\rm sep}}{d_{\rm 3D}}\right). \label{eq:condition_sphere}
\end{equation}
Assuming random orientations, this probability can be evaluated as the area fraction of the two spherical caps on a sphere of radius $d_{\rm 3D}$ that satisfy equation (\ref{eq:condition_sphere}), and the resulting probability is
\begin{equation}
    P(d_{\perp}<d_{\rm sep}\mid d_{\rm 3D}) = 1-\sqrt{1-\left(\frac{d_{\rm sep}}{d_{\rm 3D}}\right)^2}.
\end{equation}
As shown in figure\,\ref{fig:separation_redshift} (right), this probability rapidly decreases when $d_{\rm 3D}$ becomes much larger than $d_{\rm sep}$, and is particularly low for $d_{\rm 3D} \gtrsim R_{200}$.
Moreover, if the observed systems were merely projected pairs with $d_{\rm sep} \ll d_{\rm 3D}$, we would expect to find much more additional systems with larger projected separations.
This is inconsistent with the discussion in section\,\ref{s:discussion}.
Therefore, although the spectroscopic observations alone cannot provide the absolute evidence for the co-location in the same halo, the combination of small projected separations and small line-centroid offsets suggests that these systems are likely separated by less than $\sim10^2\,{\rm kpc}$, with intrinsic line-of-sight velocity offsets of $\lesssim 10^2\,{\rm km\,s^{-1}}$.

To compare with these intrinsic line-of-sight velocity offsets,
we estimate the escape velocities for a NFW-profile \citep{NFW1997} DM halo with $M_{\rm halo}=10^{11\,\mathchar`-\,11.5}M_\odot$.
Assuming an NFW potential profile,
\begin{equation}
    \Phi\left(r\right) = -\frac{G M_{\rm 200}}{r}
\frac{\ln(1+r/r_s)}{\ln(1+c_{200})-c_{200}/(1+c_{200})},
\end{equation}
where $r_s=R_{200}/c_{200}$ is the scale radius and $c_{200}$ is the concentration of the halo, the escape velocity at a separation $r$ is given by
\begin{align}
V_{\rm esc}(r)
& = \left\{-2\Phi(r) \right\}^{1/2},
\end{align}
For separations of $1\,\mathchar`-\,8\,{\rm kpc}$ and $c_{200}=3\,\mathchar`-\,4$ \citep{Diemer2015}, the escape velocity from the halo is $V_{\rm esc} \simeq 5\times10^2\,\mathchar`-\,1.1\times10^3\,{\rm km\,s^{-1}}$, depending on the assumed halo mass and concentration.
The line-of-sight velocity offset of $\Delta v_{\rm LOS, int}\lesssim10^2\,{\rm km\,s^{-1}}$ suggested above are still smaller than the estimated escape velocities for typical DM halos hosting LRDs.
Therefore, we conclude that these two LRDs are likely gravitationally bound and are unlikely to be a chance fly-by encounter system.

\subsection{Masses of central BHs}
We next discuss the masses of the SMBHs that are potentially in the process of merging.
The estimation of SMBH masses in LRDs is still under debate.
Many previous studies have applied the single-epoch method, calibrated at low redshift, 
For example, \cite{GreeneHo2005} reported the following relation using a broad H$\alpha$ line:
\begin{align}
M_{\rm BH, H\alpha} = &\left(2.0^{+0.4}_{-0.3}\right)\times10^6 \nonumber \\
&\times \left(\frac{L_{\rm H\alpha}}{10^{42}\,{\rm erg\,s^{-1}}}\right)^{0.55\pm0.02} \nonumber\\
& \times \left(\frac{{\rm FWHM}_{\rm broad,H\alpha}}{10^{3}\,{\rm km\,s^{-1}}}\right)^{2.06\pm0.06} M_\odot, \label{eq:single_epoch}
\end{align}
and these single epoch method typically infers $M_{\rm BH}\gtrsim10^7M_\odot$ for LRDs \citep[e.g.,][]{Greene2024, Maiolino2023, Harikane2023_agn, Lin2025_c3d, Matthee2024}.
If the line broadening is dominated by electron scattering not by kinematic broadening as suggested by \cite{Rusakov2025}, the inferred $M_{\rm BH}$ may be overestimated.
On the other hand, \cite{Juodzbalis2025_directmass} conducted dynamical mass measurements and reported SMBH masses comparable to those obtained from the single-epoch method.

Recently, an alternative approach has been explored based on the BH envelope model, in which the rest-optical SED is fitted with a blackbody or modified blackbody component plus Balmer absorption and/or modest dust attenuation \citep{Inayoshi2025, deGraaff2025_spec_stat, Umeda2025_BHstar, Tanaka2025_z10LRD}.
In this method, with the assumption that the BH envelope emission is entirely powered by radiation from the central BH at approximately the Eddington limit, the BH mass is inferred as
\begin{equation}
\left(\frac{M_{\rm BH, BB}}{M_\odot}\right) = \left(\frac{L_{\rm BB}}{1.26\times10^{38}\,{\rm erg\,s^{-1}}}\right) \left(\frac{\lambda_{\rm Edd}}{1.0}\right)^{-1}.\label{eq:bhe_mass}
\end{equation}
The assumption of an Eddington limit ($\lambda_{\rm Edd}\sim1$) is motivated by recent theoretical studies \citep[e.g.,][]{Kido2025, Inayoshi2025_seduni}.
They suggested that, although LRDs may accrete at super-Eddington rates, a substantial fraction of the accretion energy is transported by convection within the BH envelope and contributes to the envelope’s structural evolution, rather than being radiated away.
As a result, the effective radiative Eddington ratio is expected to take $\lambda_{\rm Edd} \sim 1$.

In this study, we estimate $M_{\rm BH}$ using both the single-epoch method (equation\,\ref{eq:single_epoch}) and the BH envelope model-based method (equation\,\ref{eq:bhe_mass}) using the SED fitting results.
Note that we apply the single-epoch method only to CW-B2-4383-\#1 and CW-A4-16093-\#1, which are covered by COSMOS-3D and have detected broad emission lines.
As summarized in table\,\ref{tab:dual_LRD_spec}, we find that the SMBH masses estimated from the broad emission lines ($\log M_{\rm BH,H\alpha}=7.1^{+0.4}_{-0.4}$ and $7.3^{+0.4}_{-0.4}$) are approximately an order of magnitude larger than those inferred from the BH envelope model fits ($\log M_{\rm BH,BB}=5.59^{+0.06}_{-0.08}$ and $6.30^{+0.04}_{-0.05}$).

Regardless of which mass estimate is more accurate, our discoveries suggest that mergers between SMBHs with $M_{\rm BH}\sim10^{6\,\mathchar`-\,7}\,M_\odot$ can occur in the early Universe.
Gravitational waves produced by mergers of such SMBHs in the early Universe will be detectable by future space-based gravitational-wave observatories such as LISA \citep{LISA2017}.
By comparing the SMBH mass functions inferred from the single-epoch method and the BH envelope model with the observed gravitational-wave detection rates, it may be possible to place independent constraints on the SMBH masses of LRDs.

An important quantity for estimating the GW event rate from LRD-LRD BH mergers is the merger rate of LRDs, which can be roughly inferred from the dual LRD fraction and the merger timescale.
If we simply estimate the dual LRD fraction using the four dual LRD candidates identified in this study and the full LRD sample size of $829$, we obtain a dual fraction of $0.5_{-0.2}^{+0.4}\%$ assuming Poisson statistics.
However, the dual fraction may depend on luminosity, and an accurate estimate requires careful consideration of sample completeness and selection biases.
Therefore, we defer a more detailed discussion of the dual LRD fraction with completeness corrections, together with the offset fraction (the fraction of LRD-galaxy merger systems) to future work.

\section{Conclusion}\label{s:consclusion}
By applying the pixel-by-pixel color selection method (figure\,\ref{fig:method_schematic}) to the COSMOS-Web imaging data, we identified the first four dual LRD candidates (figure\,\ref{fig:3col_sed}).
Through image-based modeling, we constrained the photometry and morphology of each component.
Then, we estimated their photometric redshifts from SED fitting (figure\,\ref{fig:sed_fitting}).
In addition, two of the four systems are covered by COSMOS-3D F444W grism observation, allowing us to perform line-fitting analyses.
The key results are as follows:

\begin{itemize}
    \item The SED of each component exhibits an LRD-like V-shaped profile that cannot be reproduced by brown dwarf templates (figure\,\ref{fig:sed_fitting}). Fitting with LRD models yields photometric redshifts of $z\sim5\,\mathchar`-\,7$.
    \item The projected separations between the components range from $0.\!\!^{\prime\prime}2$ to $1.\!\!^{\prime\prime}2$, corresponding to $1\,\mathchar`-\,7\,{\rm kpc}$ at their photometric or spectroscopic redshifts. Three out of the four systems, CW-B5-15958, CW-A6-19978, and CW-B2-4383 have separations of $0.\!\!^{\prime\prime}2\,\mathchar`-\,0.\!\!^{\prime\prime}4$, corresponding to physical separations of $\sim1\,\mathchar`-\,2\,{\rm kpc}$. Additionally, we detected an off-centered extended component or companion galaxy candidates around the dual LRDs.
    \item For two candidates, CW-B2-4383 (section~\ref{ss:cw-b2-4383}, figure\,\ref{fig:C3D_B2_4383}) and CW-A4-16093 (section~\ref{ss:cw-a4-16093}, figure\,\ref{fig:C3D_A4_16903}), COSMOS-3D slitless spectroscopy in F444W is available for these systems. In both systems, a single emission line is detected at the same observed wavelength in each component, strongly suggesting that the two components reside at least at the same redshift. For the brighter components, we identified a broad emission line component. For the fainter components, no broad component is detected, possibly due to the limited $S/N$. If the detected lines are H$\alpha$, the inferred redshifts are $z_{\rm H\alpha}=5.822$ and $5.464$ for CW-B2-4383 and CW-A4-16093, respectively.
    \item Based on the projected separation ($\theta$) distribution analysis of the control and mock samples, the likelihood of these candidates being chance projections of physically unrelated LRDs or BDs at different redshifts is very low (figure\,\ref{fig:separations1}). 
    \item If these candidates are indeed pairs at the same redshifts, our results suggest that LRDs show excess ($\sim20\,\mathchar`-\,50$ times) clustering on kpc scales (figure\,\ref{fig:acf}) from extrapolation of a power-law angular ACF of other LRD candidates or JWST-found spectroscopically-confirmed AGNs measured over $10^{\prime\prime}\,\mathchar`-\,600^{\prime\prime}$.
    Given that we cannot find any LRD pairs with $\theta=0\farcs4\,\mathchar`-\,4$ ($\sim2\,\mathchar`-\,20\,{\rm kpc}$) other than CW-A4-16093 while we have even three pairs with $\theta=0.\!\!^{\prime\prime}2\,\mathchar`-\,0.\!\!^{\prime\prime}4$ ($\sim1\,\mathchar`-\,2\,{\rm kpc}$), the excess clustering may indicate that these LRDs are AGNs triggered through interactions only in kpc separations.
    \item The BH masses estimated under the assumption of $\lambda_{\rm Edd}=1$ based on the BH envelope model ($M_{\rm BH,BB}=10^{5.3\,\mathchar`-\,6.3}$) are approximately one or two orders of magnitude smaller than those estimated with the single-epoch method ($M_{\rm BH,H\alpha}=10^{7.1\,\mathchar`-\,7.3}M_\odot$). Regardless of which mass estimate is more accurate, these findings suggest that mergers between SMBHs with $M_{\rm BH}\sim10^{5\,\mathchar`-\,7}M_\odot$ can occur in the early Universe. BH mergers may contribute to the early growth and evolution of SMBHs.
\end{itemize}

Two of the four candidates remain photometrically selected without spectroscopic confirmation. 
For the other two systems, only single emission lines are tentatively detected, and it remains unclear whether the fainter components exhibit broad emission lines.
Therefore, deeper spectroscopic follow-up observations are essential to confirm whether they are dual LRDs and constrain the statistics about dual LRDs (e.g., dual fractions).
Moreover, since the current sample size is small, we need to expand the sample further with large field surveys.
Investigating the environments of LRDs with larger spectroscopic samples and comparing them to other AGNs and galaxies will also be critical for understanding their nature.

\begin{ack}
We thank Khee-Gan Lee and Wout Goesaert for the fruitful discussion.
We thank the anonymous referee for helpful feedback.
This work is based on observations made with the NASA/ESA/CSA James Webb Space Telescope.
The data were obtained from the Mikulski Archive for Space Telescopes at the Space Telescope Science Institute, which is operated by the Association of Universities for Research in Astronomy, Inc., under NASA contract NAS 5-03127 for JWST.
These observations are associated with program IDs 1727 and 5893.
Numerical computations were in part carried out on the iDark and GPGPU cluster, Kavli IPMU.
This work was made possible by utilising the CANDIDE cluster at the Institut d’Astrophysique de Paris. The cluster was funded through grants from the PNCG, CNES, DIM-ACAV, the Euclid Consortium, and the Danish National Research Foundation Cosmic Dawn Center (DNRF140). It is maintained by Stephane Rouberol.
\end{ack}

\section*{Funding}
Kavli IPMU is supported by World Premier International Research Center Initiative (WPI), MEXT, Japan.
TST is supported by Japan Society for the Promotion of Science (JSPS) KAKENHI Grant Number JP25KJ0750 and the Forefront Physics and Mathematics Program to Drive Transformation (FoPM), a World-leading Innovative Graduate Study (WINGS) Program at the University of Tokyo.
JDS is supported by JSPS KAKENHI (JP22H01262).
JA is supported by the JSPS KAKENHI Grant Number JP24KJ0858.
KI acknowledges support from the National Natural Science Foundation of China (12073003, 12003003, 11721303, 11991052, 11950410493), and the China Manned Space Project (CMSCSST-2021-A04 and CMS-CSST-2021-A06).
MA is supported by the Data-Scientist-Type Researcher Training Project of The Graduate University for Advanced Studies, SOKENDAI.
J-TS acknowledges funding by the Deutsche Forschungsgemeinschaft (DFG, German Research Foundation) - Project number 518006966 (J.-T.S.) and support by the DFG under Germany’s Excellence Strategy - EXC 2121 „Quantum Universe“ - 390833306.
RAM acknowledges support from the Swiss National Science Foundation (SNSF) through project grant 200020\_207349.


\appendix 
\section{Search for dual LRD candidates in other imaging survey fields}\label{ap:others}
\begin{figure*}
 \begin{center}
  \includegraphics[width=17.8cm]{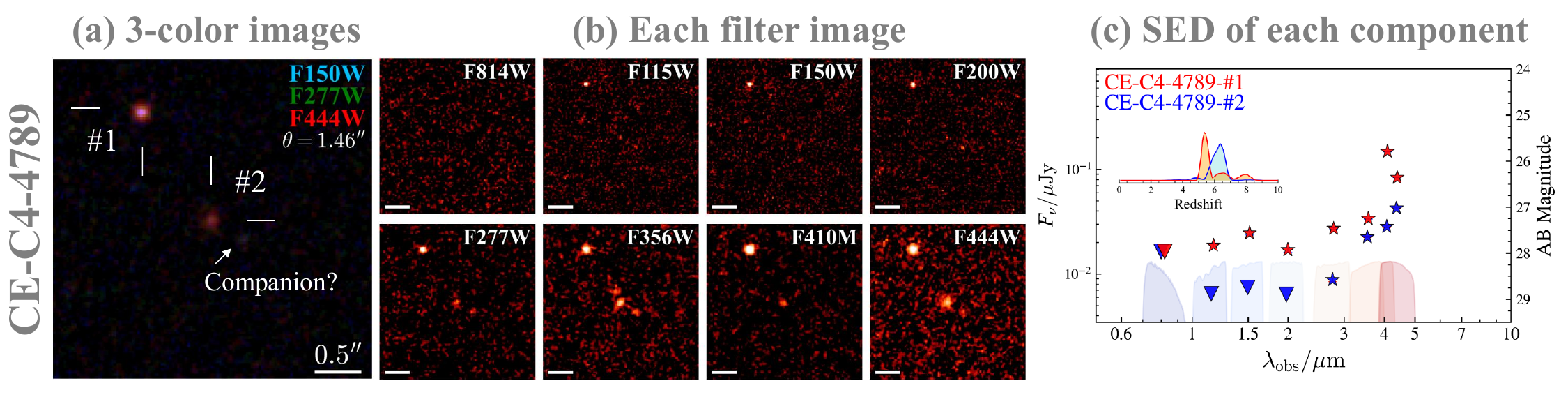} 
 \end{center}
\caption{
Same as figure\,\ref{fig:3col_sed}, but for CE-C4-4789
}
\label{fig:3col_sed_app}
\end{figure*}

As described in section\,\ref{ss:other_fields}, we also apply the same selection method to the other JWST fields, CEERS, JADES, and PRIMER-UDS.
Through visual inspection after the pixel-by-pixel color selection, we identified one system in the CEERS field, CE-C4-4789 (figure\,\ref{fig:3col_sed_app} and table\,\ref{tab:dual_LRD_app}), that exhibits similar appearance to a dual LRD.
In this system, component\,\#2 has more than 10 contiguous pixels satisfying the LRD-like color criterion ($m_{\rm F277W}-m_{\rm F444W} > 1.5$), whereas component\,\#1, separated by $1.\!\!^{\prime\prime}46$, does not meet our threshold of at least 10 pixels fulfilling this color requirement.
The measured $m_{{\rm F277W}} - m_{\rm F444W}$ color is 1.83 for component\,\#2, while component\,\#1 shows a significantly bluer color of 1.27.
Consequently, component\,\#1 fails to satisfy our adopted color threshold, and the system is not classified as a dual LRD according to the definition used in this study.
We also note that although component\,\#2 is identified as an LRD in \cite{Kocevski2024}, component\,\#1 is not selected, consistent with our classification.
Furthermore, component\,\#1 shows a clear flux excess in F410M (figure\,\ref{fig:3col_sed_app} c), indicative of strong H$\alpha$ emission, whereas component\,\#2 exhibits no such excess.
This discrepancy raises the possibility that the two components are not at the same redshift, since LRDs typically exhibit strong H$\alpha$ emission that can significantly contribute even to broadband photometry.
Such a possible redshift difference is also consistent with the estimated photometric redshifts (figure\,\ref{fig:3col_sed_app} c and table\,\ref{tab:dual_LRD_app}).
Spectroscopic observations are required to determine the nature of this system.

\begin{table}[]
\centering
\begin{threeparttable}
\caption{Characteristics of CE-C4-4789.
}\label{tab:dual_LRD_app}
\begin{tabular}{c | ll}
& \multicolumn{2}{c}{CE-C4-4789} \\
& \#1 & \#2 \\
\hline\hline
R.A. & \timeform{14h18m48s.451} & \timeform{14h18m48s.404} \\
Decl. & \timeform{52D44'38''.69} & \timeform{52D44'40''.09} \\
$F_{\rm F814W}/{\rm nJy}$\tnote{a} & $<16$ & $<16$  \\
$F_{\rm F115W}/{\rm nJy}$\tnote{a} & $18.7\pm0.9$ & $<6.5$ \\
$F_{\rm F150W}/{\rm nJy}$\tnote{a} & $24.6\pm1.0$ & $<7.4$ \\
$F_{\rm F200W}/{\rm nJy}$\tnote{a} & $17.0\pm0.8$ & $<6.3$ \\
$F_{\rm F277W}/{\rm nJy}$ & $27.2\pm0.4$ & $8.8\pm0.4$ \\
$F_{\rm F356W}/{\rm nJy}$ & $33.8\pm0.4$ & $22.6\pm0.4$ \\
$F_{\rm F410M}/{\rm nJy}$ & $148.2\pm0.8$ & $28.5\pm0.7$ \\
$F_{\rm F444W}/{\rm nJy}$ & $83.6\pm0.6$ & $42.5\pm0.6$ \\
$\theta_{\rm F277W}/{\rm arcsec}$ & \multicolumn{2}{c}{1.46} \\
\hline
$z_{\rm photo}$ & $5.43_{-0.16}^{+1.35}$ & $6.21_{-0.47}^{+0.31}$ \\
$T_{\rm eff}/{\rm K}$ & $\left(3.2^{+1.3}_{-0.8}\right)\times10^3$ & $\left(6.0^{+1.7}_{-1.7}\right)\times10^3$ \\
$\log\left(L_{\rm BB}/{\rm erg\,s^{-1}}\right)$ & $43.23_{-0.73}^{+0.95}$ & $43.13_{-0.17}^{+0.07}$ \\
$\log\left(M_{\rm BH, BB}/M_\odot\right)$\tnote{b} & $5.13_{-0.73}^{+0.95}$ & $5.03_{-0.17}^{+0.07}$ \\
\hline
\end{tabular}
\begin{tablenotes}
\footnotesize
\item[a] Upper limits are based on the $3\sigma$ noise level.
\item[b] Black hole masses estimated from the fitted blackbody luminosity assuming Eddington-limited accretion.
\end{tablenotes}
\end{threeparttable}
\end{table}

\section{Non-detection of broad line due to faintness}\label{ap:broad_line}
For components \#2 in both CW-B2-4383 and CW-A4-16093, single-Gaussian fitting results in a lower BIC than double-Gaussian fitting.
Especially in CW-A4-16093, FWHM from the single Gaussian fitting is ${\rm FWHM_{single}} =348^{+96}_{-77}\,{\rm km\,s^{-1}}$, indicating that a broad H$\alpha$ component is not detected (figure\,\ref{fig:C3D_A4_16903}).
However, given the low $S/N$, it remains unclear whether the lack of detected broad emission in \#2 arises from the intrinsic absence of a broad line or merely from insufficient sensitivity.
Therefore, in this appendix section, we explore potential scenarios under which a broad line would exist but remain undetectable through mock analysis. 

\begin{figure}
 \begin{center}
  \includegraphics[width=8cm]{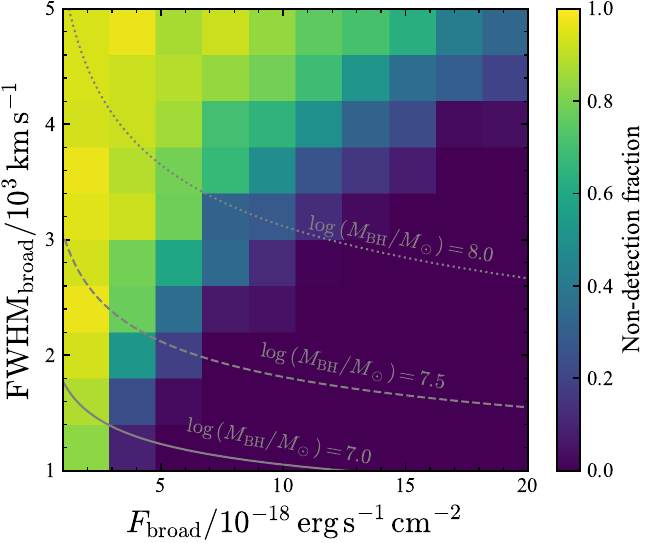}
 \end{center}
\caption{
Distribution of the fraction of mock spectra in which the broad component cannot be detected, while the line flux from single-Gaussian fitting is consistent with the value measured from the real spectrum of CW-A4-16093 component\,\#2, shown as a function of broad-line FWHM and flux.
The gray curves correspond to constant BH masses of $\log\left(M_{\rm BH, H\alpha}\right)=7.0$, $7.5$, and $8.0$, when assuming the single-epoch BH mass estimator (equation\,\ref{eq:single_epoch}).
If component\,\#2 of CW-A4-16093 has a relatively low broad-line flux or a very large line width, the broad component cannot be reliably detected in COSMOS-3D-like spectra.
}
\label{fig:bl_det}
\end{figure}

In the mock data generation, assuming the case of component\,\#2 in CW-A4-16093, we fix the narrow-line FWHM to $350\,{\rm km\,s^{-1}}$ and the narrow-line flux to $5\times10^{-18}\,{\rm erg\,s^{-1}\,cm^{-2}}$.
We then vary the broad-line FWHM in $[1000, 5000]\,{\rm km\,s^{-1}}$ and broad-line flux in $[1\times10^{-18}, 20\times10^{-18}]\,{\rm erg\,s^{-1}\,cm^{-2}}$, and generate mock spectra with noise levels comparable to those of the real COSMOS-3D spectrum.
These mock spectra are fitted with both single-Gaussian and double-Gaussian models with the same way as section\,\ref{sss:line_fit}.
For each combination of broad-line parameters, we perform 100 realizations of the data generation and fittings.
Figure\,\ref{fig:bl_det} summarizes the Monte Carlo-derived probability that the broad component remains undetected from the BIC comparison between single and double Gaussian fittings while the line flux estimated from single-Gaussian fitting matches the observed flux range ($3\times10^{-18} < F_{\rm single}/{\rm erg\,s^{-1}\,cm^{-2}} < 6\times10^{-18}$).
Our mock analysis suggests that broad components with relatively low flux ($F_{\rm broad}\lesssim5\times10^{-18}\,{\rm erg\,s^{-1}\,cm^{-2}}$) or large line widths (${\rm FWHM_{broad}}\gtrsim 3000\,{\rm km\,s^{-1}}$) are still consistent with the non-detection of a broad line in the current COSMOS-3D spectrum.
Therefore, the existing C3D observations alone cannot rule out the presence of a broad component.
Deeper spectroscopic observations are required to robustly determine whether a broad line is present.
Note that if we assume an intrinsic broad-line FWHM of ${\rm FWHM_{broad}}< 3000\,{\rm km\,s^{-1}}$, we can further constrain the intrinsic broad-line flux to be $F_{\rm broad} \lesssim 1\times10^{-17}{\rm erg\,s^{-1}\,cm^{-2}}$ ($L_{\rm broad} \lesssim 3\times10^{-42}{\rm erg\,s^{-1}}$ at the same redshift as CW-A4-16093). 
The corresponding upper limit for the total line EW for CW-A4-16093 component \#2 is $< 7\times10^2\,{\text{\AA}}$, which remains consistent with the photometrically constrained EW derived from the SED fitting.

\bibliographystyle{plainnat2}
\bibliography{export}{}

\end{document}